\documentclass{article}

    \usepackage[preprint]{neurips_2025}

\usepackage[utf8]{inputenc} 
\usepackage[T1]{fontenc}    
\usepackage{hyperref}       
\usepackage{url}            
\usepackage{booktabs}       
\usepackage{amsfonts}       
\usepackage{nicefrac}       
\usepackage{microtype}      
\usepackage{xcolor}         
\usepackage{wrapfig}

\usepackage[pdftex]{graphicx}

\usepackage{xspace}
\usepackage{xcolor}
\usepackage{tcolorbox}
\usepackage{multirow}

\usepackage{pifont}
\newcommand{\xmark}{\ding{55}}%
\newcommand{\cmark}{\ding{51}}%

\newcommand{\ie}{\textit{i.e.,} }

\newcommand{\eg}{\textit{e.g.,} }

\newcommand{\sys}{\textsc{WilliamT}\xspace}

\definecolor{myorange}{HTML}{fce5cd}
\definecolor{myblue}{HTML}{cfe2f3}
\newcommand{\orangesquare}{\textcolor{myorange}{\rule{1.5ex}{1.5ex}}}
\newcommand{\bluesquare}{\textcolor{myblue}{\rule{1.5ex}{1.5ex}}}

\definecolor{pinkshocking}{rgb}{0.99,0.01,0.84}

\usepackage{listings}
\usepackage{subcaption}

\definecolor{codegreen}{rgb}{0,0.6,0}
\definecolor{codegray}{rgb}{0.5,0.5,0.5}
\definecolor{codepurple}{rgb}{0.58,0,0.82}
\definecolor{backcolour}{rgb}{0.95,0.95,0.92}

\lstdefinestyle{mystyle}{
    backgroundcolor=\color{backcolour},
    commentstyle=\color{codegreen},
    keywordstyle=\color{magenta},
    numberstyle=\tiny\color{codegray},
    stringstyle=\color{codepurple},
    basicstyle=\ttfamily\footnotesize,
    breakatwhitespace=false,
    breaklines=true,
    captionpos=b,
    keepspaces=true,
    numbers=left,
    numbersep=5pt,
    showspaces=false,
    showstringspaces=false,
    showtabs=false,
    tabsize=2
}
\lstdefinestyle{cpp}{
    language=C++,
    basicstyle=\ttfamily,
    keywordstyle=\color{blue}\ttfamily,
    stringstyle=\color{red}\ttfamily,
    commentstyle=\color{codegreen}\ttfamily,
    morecomment=[l][\color{magenta}]{\#}
}

\definecolor{lightgray}{gray}{0.95}

\lstdefinestyle{stacktrace}{
  backgroundcolor=\color{lightgray},
  basicstyle=\ttfamily\small,
  breaklines=true,
  frame=single,
  columns=fullflexible,
  showstringspaces=false
}

\title{Fixing 7,400 Bugs for 1\$: \\ Cheap Crash-Site Program Repair}

\author{%
  Han Zheng \\
  EPFL\\
  Lausanne, Switzerland \\
  \texttt{han.zheng@epfl.ch} \\
  \And
  Ilia Shumailov \\
  Google DeepMind \\
  London, UK \\
  \texttt{iliashumailov@google.com} \\
  \AND
  Tianqi Fan \\
  Google \\
  Zurich, Switzerland \\
  \texttt{tqfan@google.com} \\
  \And
  Aiden Hall \\
  Google \\
  New York, USA \\
  \texttt{aidenhall@google.com} \\
  \And
  Mathias Payer \\
  EPFL \\
  Lausanne, Switzerland \\
  \texttt{mathias.payer@nebelwelt.net} \\
}

\begin{document}

\maketitle

\begin{abstract}

    The rapid advancement of bug-finding techniques has led to the 
    discovery of more vulnerabilities than developers can reasonably fix, 
    creating an urgent need for effective Automated Program 
    Repair (APR) methods. However, the complexity of modern bugs often 
    makes precise root cause analysis difficult and unreliable.
    To address this challenge, we propose \emph{crash-site repair} to 
    simplify the repair task while still mitigating the risk of 
    exploitation. In addition, we introduce a \emph{template-guided patch 
    generation} approach that significantly reduces the token cost of 
    Large Language Models (LLMs) while maintaining both efficiency and 
    effectiveness.
    
    We implement our prototype system, \sys, and evaluate it against 
    state-of-the-art APR tools. Our results show that, when combined 
    with the top-performing agent CodeRover-S, \sys reduces token cost 
    by 45.9\% and increases the bug-fixing rate to 73.5\% (+29.6\%) 
    on ARVO, a ground-truth open source software vulnerabilities benchmark.
    Furthermore, 
    we demonstrate that \sys can function effectively even without 
    access to frontier LLMs: even a local model running on a Mac M4 Mini 
    achieves a reasonable repair rate. These findings highlight 
    the broad applicability and scalability of \sys.

\end{abstract}

\section{Introduction}
\label{sec:intro}

Modern software systems are extremely complex. While this ever-increasing complexity enables richer functionality, it simultaneously introduces a significant number of 
(exploitable) vulnerabilities~\cite{chrome_exploit_bh_24_1,
chrome_exploit_bh_24_2,chrome_exploit_bh_23_1,chrome_exploit_bh_22_1,
chrome_exploit_bh_22_2,kernel_exploit_bh_22,macos_exploit_bh_21}.
To proactively safeguard users from malicious threats, fuzzing~\cite{afl,
fioraldi2020afl++,libfuzzer,zheng2023fishfuzz,zheng2025mendelfuzz} 
emerged to automate the bug finding. Over the years fuzzing demonstrated its extreme usefulness and efficiency even in the most complex and security-critical 
systems, including but not limited to web browsers~\cite{xu2020freedom,zhou2022minerva,
gross2023fuzzilli,wachterdumpling}, Linux kernels~\cite{syzkaller,
fleischer2023actor}, MacOS~\cite{zhu2024crossfire,yin2023kextfuzz}, 
Hypervisors~\cite{pan2021v,liu2023videzzo,bulekov2024hyperpill,matruman} 
and Trust Execution Environments~\cite{wang2024syztrust}.
However, the advancements in fuzzing also 
increase demands on developers tasked with fixing the discovered vulnerabilities. For example, despite extensive 
maintenance efforts, more than 1,500 kernel bug reports remain 
unresolved on Syzbot~\cite{syzbot_open_bugs}. Modern fuzzers uncover 
vulnerabilities at a pace that far outstrips the ability of developers 
to address them. What is worse, the process of fixing bugs requires renewed fuzzing efforts to check the fixes, which in turn often reveals additional bugs. 

This growing discrepancy between the bug discovery and bug fix rate 
has led to a backlog that 
exceeds developers’ capacity to triage and remediate 
effectively~\cite{too_many_bugs_fix}. These challenges underscore the 
urgent need for Automated Program Repair (APR) solutions. Even imperfect, 
such solutions would significantly improve the security of our systems 
and allow the developers to focus on the most complex of bugs. 
Recent advancements in Large 
Language Models (LLMs)~\cite{team2024gemma,liu2024deepseek,guo2025deepseek,
grok,chatgpt} offer a promising direction for addressing this challenge. 
Leveraging the capabilities of LLMs, a variety of novel Automatic 
Program Repair (APR) techniques have emerged~\cite{zhang2024fixing,
huang2025template,huang2023empirical,tang2024code,zhang2024autocoderover}.

Despite their promise, current APR agents struggle as they attempt to address bugs 
by identifying and fixing their root causes. This strategy
is often impractical -- root cause analysis is inherently complex, if 
not impossible, and typically requires expensive, heavyweight 
LLM backends, making it 
difficult for individual developers to adopt and deploy these solutions 
at scale. Moreover, root cause analysis relies on the LLMs’ ability to 
accurately perform multiple reasoning tasks~\cite{zhang2024autocoderover,
zhang2024fixing}, and imprecision in any of these stages can necessitate 
repeated attempts at repair~\cite{xia2024automated}, further increasing 
the number of LLM queries and the associated runtime costs.

In contrast to these imprecise, high-cost root-cause-oriented methods, in this paper 
we advocate for an alternative lightweight, ad-hoc repair approach that 
prevents immediate exploitation~\cite{chrome_checklist} rather than fixing 
the bug at the root cause. 
Root-cause fixes often require a deep understanding of the program and 
complex code changes in components that may not appear in the crash 
stack, whereas crash-site fixes are quicker and can block exploitation 
by adding a simple assertion right before crash site.
This thereby grants developers more time for an in-depth investigation while the immediate 
threat is contained. To this end, we 
propose \sys, a new template-based APR agent specifically designed to 
repair memory corruption vulnerabilities given a Proof-of-Concept (PoC) 
input.
\sys is inspired by the folk hero William Tell who had, under pressure, 
one shot to hit an apple on his kid's head. Similarly, \sys aims for 
"one-shot" fixes to promptly block exploitation.
\sys operates by taking fuzzer-generated PoCs and Sanitizer 
outputs~\cite{serebryany2012addresssanitizer} as input. It employs a 
regular-expression-based method for both fault localization and patch 
generation, reserving LLM usage exclusively for root cause analysis when 
necessary. By minimizing reliance on LLMs and constraining their outputs 
via predefined bug templates, \sys substantially reduces query costs and 
enables deployment with smaller and cheaper LLMs. This makes \sys a 
scalable and practical APR solution, particularly suitable for 
resource-constrained environments.

We leverage ARVO~\cite{mei2024arvo}, a ground-truth open source 
software vulnerability benchmark for our evaluation and compare 
\sys against state-of-the-art APR agents based on LLMs. 
While the CodeRover-S~\cite{zhang2024fixing} is the best-performing 
agent, \sys reduces its token usage by 99.7\% while retaining over 
86.7\% of the CodeRover-S. 
Moreover, when combined with CodeRover-S, 
\ie fixing using \sys and forward the unfixed bugs to CodeRover-S, 
the optimized pipeline reduces token cost by 45.9\% and improves 
the fixing rate by 29.6\% compared to CodeRover-S alone.
We further extend \sys to work with both cutting-edge and local LLMs. 
Our findings indicate that \sys not only integrates effectively with 
frontier LLMs but also scales efficiently when deployed with local 
models running on a Mac Mini, demonstrating its broad applicability 
and scalability.
To sum up, our contributions are:
\begin{itemize}
    \item We propose \emph{crash-site repair} to avoid 
    complex and imprecise root cause analysis.
    \item We introduce a \emph{template-guided patch generation} technique 
    to reduce LLM inference cost.
    \item We implement and evaluate \sys, demonstrating that—when 
    combined with a leading SoTA method—it achieves a 45.9\% cost 
    reduction and a 29.6\% increase in fixing precision.
    \item We promise to fully release \sys upon paper acceptance to 
    support open science.
\end{itemize}

\section{Background}
\label{sec:background}

\subsection{LLM-Based Program Repairing}
\label{ssec:llm-apr}

\begin{figure*}[t]
	\centering
	\includegraphics[width=0.95\linewidth]{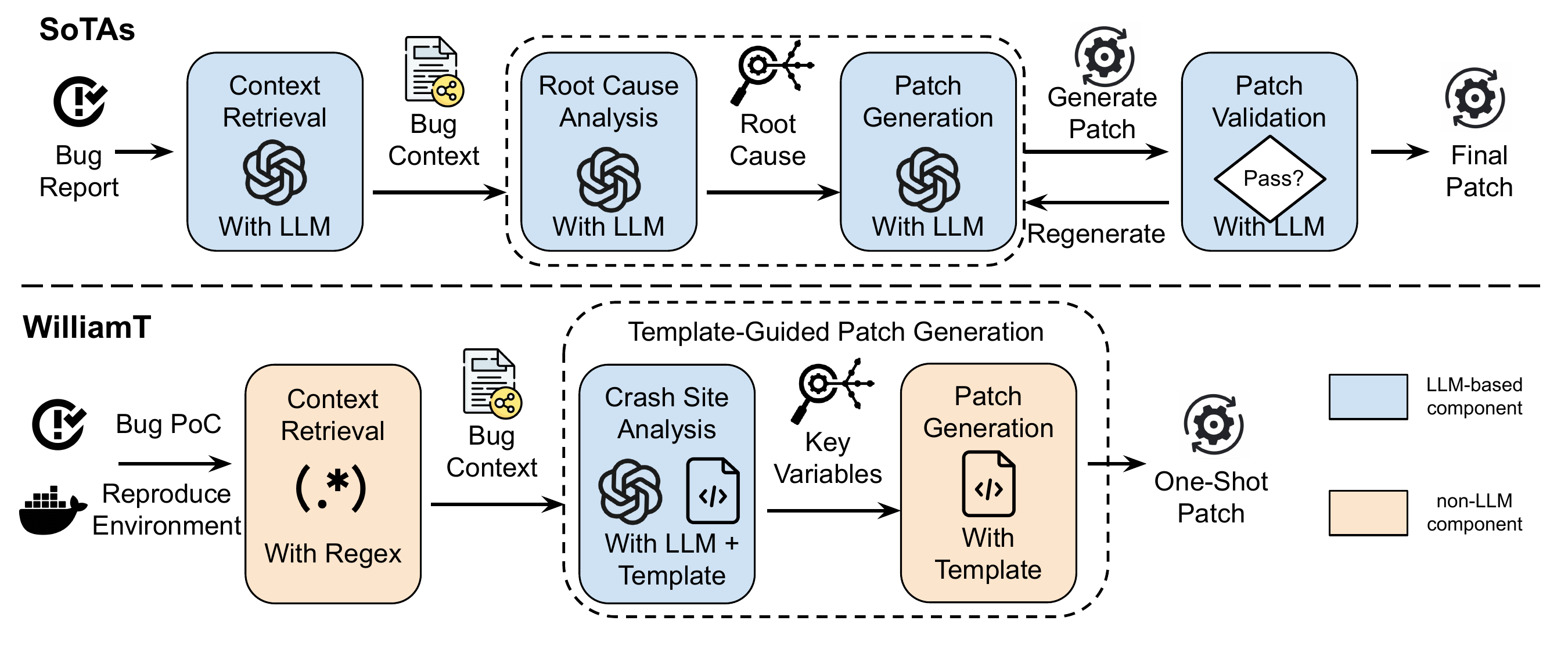}
	\caption{Workflow of Current LLM-based Program Repairing Agents vs \sys.
    \bluesquare{} stands for the LLM components while \orangesquare{} means non-LLM modules.}
	\label{fig:workflow-llm-apr}
\end{figure*}

The emergence of Large Language Models (LLMs) has inspired 
researchers to explore their potential in enabling agents for Automatic 
Program Repair (APR). In contrast to earlier approaches that relied on 
supervised model training or LLM fine-tuning~\cite{tufano2019empirical, 
chen2019sequencer,feng2020codebert,huang2025template}, agent-based 
methods benefit directly from the continual advancements in foundational 
LLMs~\cite{llama4, gemma3, liu2024deepseek, guo2025deepseek, grok, 
qwq32b, qwen2.5}, without necessitating additional manual tuning.

To further reduce the pretraining and fine-tuning costs, recent efforts 
have proposed agent-based APR frameworks~\cite{xia2022less,xia2024automated}. 
These frameworks typically consist of three core components, as depicted 
in~\autoref{fig:workflow-llm-apr}. First, the agent processes the bug 
report, retrieves relevant context, and submits it to the LLM for 
root cause analysis and patch generation. The generated patch is then 
passed back to the LLM for validation. If the patch fails validation, the process iterates, 
returning to the patch generation phase, until a satisfactory patch 
is produced.

However, current agents delegate all four steps—retrieval, root cause 
analysis, patch generation, and validation—entirely 
to LLMs. This results in a substantial increase in token consumption. 
Moreover, the quality of each module's output is highly dependent on 
the size and capacity of the underlying LLM, making it impractical to 
deploy such agents using lightweight or locally hosted models.
This challenge highlights the need for a more efficient agent 
architecture that can scale with smaller LLMs, while minimizing reliance 
on the model's capacity and reducing overall token consumption.

\subsection{Memory Corruption Vulnerabilities and OSS-Fuzz Benchmarks}
\label{ssec:apr-mem-bench}

Over the past 30 years, memory corruption vulnerabilities remain critical 
in modern software systems~\cite{hack_stack1996,memory_safety_chrome,
memory_safety_android}. This persistence underscores the importance of 
APR tools and their ability to address such vulnerabilities. 
Consequently, the capability to fix memory corruption bugs has 
become a key metric for evaluating APR tools.

Memory corruption vulnerabilities can be broadly divided into two major 
categories~\cite{Payer18SS3P}: spatial and temporal memory corruptions 
(\autoref{list:memory-corruption-examples}). 
\textit{Spatial memory corruption} occurs when a program accesses memory outside 
the bounds of an allocated buffer. These accesses can involve the stack, 
heap, or global memory regions.
\textit{Temporal memory corruption}, on the other hand, results from 
accessing a buffer that has been deallocated or is no longer valid at the time of 
access. This includes accessing an already-freed heap buffer or a 
returned stack buffer that is no longer valid.

\begin{figure}[t!]
    \begin{subfigure}{0.49\linewidth}  
    \centering
    \begin{lstlisting}[style=cpp]
// stack/heap/global buffer
type *buf = get_buffer(SIZE); 
// idx < 0 or >= SIZE
type value = buf[idx]; \end{lstlisting}
    \caption{Spatial Memory Corruption}
    \end{subfigure}
    \hfill
    \begin{subfigure}{0.49\linewidth}  
    \centering
    \begin{lstlisting}[style=cpp]
// buffer is invalid
invalid_buffer(buf);
// reuse the content in buffer
type value = buf[idx];
\end{lstlisting}
    \caption{Temporal Memory Corruption}
    \end{subfigure}
    \caption{Example Memory Corruption Vulnerabilties.}
    \label{list:memory-corruption-examples}
\end{figure}

Memory corruption bugs are prevalent in open-source software and often 
propagate into security-critical systems such as Chrome and 
iOS~\cite{chrome_third_party_bug,apple_third_party_bug}. To mitigate 
these risks, Google introduced OSS-Fuzz~\cite{oss_fuzz}, a continuous fuzzing 
infrastructure designed to test open-source software (OSS) at scale and 
protect users against potential vulnerabilities. 
As of now, OSS-Fuzz has discovered and helped fix over 36,000 bugs—
many of which were memory-related~\cite{oss_fuzz_blog}.
This rich and diverse dataset makes OSS-Fuzz an ideal benchmark 
for evaluating APR tools. ARVO~\cite{mei2024arvo}, inspired by OSS-Fuzz, 
reconstructs and curates a reproducible dataset of OSS bugs specifically 
tailored for APR evaluation. Compared to other ground-truth memory 
corruption datasets~\cite{gao2021beyond,fan2020ac,bhandari2021cvefixes}, 
ARVO automates the entire compilation pipeline and ensures 
reproducibility. This automation facilitates fair and consistent 
comparisons across different APR tools.

\section{Primer on Bug Fixing}
\label{sec:motivation}

\begin{figure}[t!] 
    \begin{lstlisting}[style=cpp]
// [1] crash site: yy_c >= sizeof(yy_current_state)
// /src/igraph/build/src/io/parsers/dl-lexer.c:3536
(yy_trans_info = &yy_current_state[yy_c])->yy_verify == yy_c;

// [2] root cause: wrong compile flag
// /src/CMakeLists.txt
- COMPILE_FLAGS "${flex_hide_line_numbers} -CF -8"
+ COMPILE_FLAGS "${flex_hide_line_numbers} -Cf -8"

\end{lstlisting}
\caption{Root Cause and Crash Site of Bug 66992.}
\label{list:motivation-example}
\end{figure}

The objective of automated program repair (APR) tools can be broadly 
categorized into three progressive stages: gracefully crashing, 
bailing out and aborting the function, and fixing the root caus of 
the bug.
In the first stage, the primary goal is to prevent potential 
exploitation. Patches at this level may terminate program execution 
immediately upon detecting a fault, thereby avoiding the execution of 
potentially dangerous code paths.
In the second stage, patches aim to handle the error more gracefully 
by returning appropriate error codes or statuses to the caller. 
This ensures the program remains operational and that the presence of 
a bug does not compromise overall system availability.
Finally, the most desirable stage involves addressing the root cause 
of the bug, thereby preventing the fault from occurring in the first 
place. This level of repair reflects a complete and semantically 
correct fix that preserves both safety and functionality.

While many agent-based APR tools aim to address 
vulnerabilities at their root cause, this is not always practical. 
In real-world scenarios, APR systems typically operate on limited 
inputs—either a natural language bug report or a sanitizer-generated 
crash report—which only provide information about the \textit{crash site}. 
However, the \textit{root cause} may reside in an entirely different 
source file or even in a separate component of the system.
For instance, in the case illustrated in \autoref{list:motivation-example}, 
a global buffer overflow vulnerability arises in the \textit{igraph} library due 
to the index variable \lstinline{yy_c} exceeding the bounds of the 
global array \lstinline{yy_current_state}. Importantly, this code was 
not manually written but automatically generated by \lstinline{flex}. 
The true root cause lies in the incorrect use of compilation flags, 
which the developer ultimately resolved by modifying the CMakefile. 
This example demonstrates that effective repair sometimes requires 
cross-domain analysis across build systems, code generators, and 
runtime behaviors—capabilities that are beyond the scope of current 
large language models (LLMs), due to limitations in contextual capacity 
and abstract reasoning.

To address this challenge, we draw inspiration from 
Chrome’s development guide~\cite{chrome_checklist}, which recommends 
inserting CHECK statements (\ie assertions) to mitigate the risk of 
exploitation before a root cause fix is available. This strategy 
transforms potentially exploitable vulnerabilities into intentional, 
controlled crashes, giving developers time to implement a permanent 
solution.

Building on this principle, \sys directly inserts the patches before 
crash site by leveraging the predefined templates and \emph{crash site analysis} 
(as illustrated in ~\autoref{sec:append-fix-template}). 
Specifically, we 
focus on four widely prevalent categories of exploitable 
vulnerabilities~\cite{mei2024arvo}: Heap-Buffer-Overflow (HBO), 
Global-Buffer-Overflow (GBO), Stack-Buffer-Overflow (SBO), and 
Heap-Use-After-Free (UAF). Each category is associated with a 
template that encapsulates common crash patterns and patch 
strategies.
This \emph{crash-site repair} paradigm significantly simplifies the 
complexity of bug fixing and circumvents the need for precise root 
cause localization. As a result, \sys achieves comparable repair 
effectiveness while requiring fewer than 1\% of the LLM queries 
typically consumed by root-cause-based systems, making it more 
scalable to resource-constrained or local LLM environments.

\section{\sys Design}
\label{sec:design}

We propose \sys, a lightweight and scalable system designed to 
eliminate the excessive query cost associated with LLMs,  
scaling APR agents to smaller model.
\sys takes as input a ClusterFuzz-generated Proof-of-Concept 
(PoC)~\cite{clusterfuzz} and a reproducible Docker image as input, 
finally produces a one-shot patch for the vulnerable program. 
\sys comprises two main components: \textbf{Regex-Based Context Retrieval} 
and \textbf{Template-Guided Patch Generation}, as illustrated in 
\autoref{fig:workflow-llm-apr}.

\noindent \textbf{Regex-Based Context Retrieval.} 
By reproducing the PoC within the provided Docker environment, \sys 
utilizes the sanitizer report~\cite{serebryany2012addresssanitizer} to 
identify the crash site through regex-based matching. This approach 
offloads the context extraction task from token-intensive LLM queries, 
significantly improving performance and scalability. While the crash 
site is not always equivalent to the true root cause, generating a 
patch just before the crash point is sufficient to mitigate 
exploitation~\cite{chrome_checklist}. 
Therefore, \sys provides the LLM agent with a focused 
context surrounding the crash site.
We include technical implementations in~\autoref{sec:append-regex-prompt}.

\noindent \textbf{Template-Guided Patch Generation.}
Directly prompting an LLM to generate a fix often results in arbitrary 
and inconsistent outputs, highly dependent on the LLM’s internal 
capabilities. To address this, \sys categorizes each bug based on its 
memory corruption type (\eg, heap-buffer-overflow, 
global-buffer-overflow, stack-buffer-overflow, heap-use-after-free). 
For each bug category, \sys selects a predefined patch template and 
guides the LLM to identify only the key variables relevant to the 
patch. This template-based constraint significantly reduces the 
LLM’s cognitive burden by offloading most of the patch generation 
logic, requiring the LLM only to perform minimal crash  
site analysis.
We provide concrete details in~\autoref{sec:append-fix-template}. 
This agent design enables 
smaller LLMs to achieve reasonable performance with 
significantly reduced resource requirements.

\section{Evaluation}
\label{sec:evaluation}


We evaluate \sys to answer the following research questions (RQs):
1) RQ1: Does \sys outperforms other SoTAs in terms of bug fixing 
capability? 
2) RQ2: Does \sys scale on smaller LLMs? 
3) RQ3: What kind of bugs does \sys and SoTAs fix? 

\noindent \textbf{Benchmark Selection.} 
We utilize ARVO~\cite{mei2024arvo}, a high-quality dataset comprising 
over 5,000 ground-truth memory corruption bugs drawn from more than 
250 real-world software projects. In contrast to manually constructed 
benchmarks~\cite{cgc_binary} or non-reproducible 
datasets~\cite{bhandari2021cvefixes,fan2020ac}, ARVO ensures that all 
bugs are both ground-truth and reproducible. Moreover, each bug in 
ARVO originates from OSS-Fuzz~\cite{oss_fuzz}, thereby closely aligning 
the dataset with real-world vulnerability discovery scenarios.
Specifically, we choose all HOF, SOF, UAF and GOF bugs (358 bugs) 
that can be compiled within 15 minutes following the 
recommended practice~\cite{zhang2024fixing}.

\noindent \textbf{Repair Metrics.} We classify the 
fix into following categories~\cite{huang2023empirical,zhang2024fixing}: 
No Code: the code is generated by build system; 
No Patch: LLM failed to generate patch or the  
patched program does not compile; Implausible: patched 
program compiles but still crashes when running with PoC; 
Plausible: patched program does not crash when taking the PoC 
as input. We consider the plausible rate as the successful 
fixing rate to align with other SoTAs~\cite{zhang2024fixing}.

\noindent \textbf{Evaluation Setup.} All our agents are running on a 
Ubuntu 22.04 server equipped with AMD EPYC 7302P and 64GB RAM.
We use the official APIs to access the frontier LLM models 
and run local LLMs on a RTX 4090 GPU. 
We also test the local model Gemma3:4b on Mac Mini M4 (16 GB RAM) 
and it performs on par to RTX 4090, illustrating \sys's scalability.

\noindent \textbf{LLM Selection.} 
We select DeepSeek (V3, R1), ChatGPT (4o, o3-mini), 
Claude (3.5-Haiku, 3.7-Sonnet) as frontier LLM baselines, 
and choose Gemma3 (27B, 12B, 4B, 1B) as local LLMs.

\subsection{RQ1: Does \sys Outperforms Other SoTAs?}
\label{ssec:eval-rq1}

\begin{figure*}[ht]
	\centering
	\includegraphics[width=\linewidth]{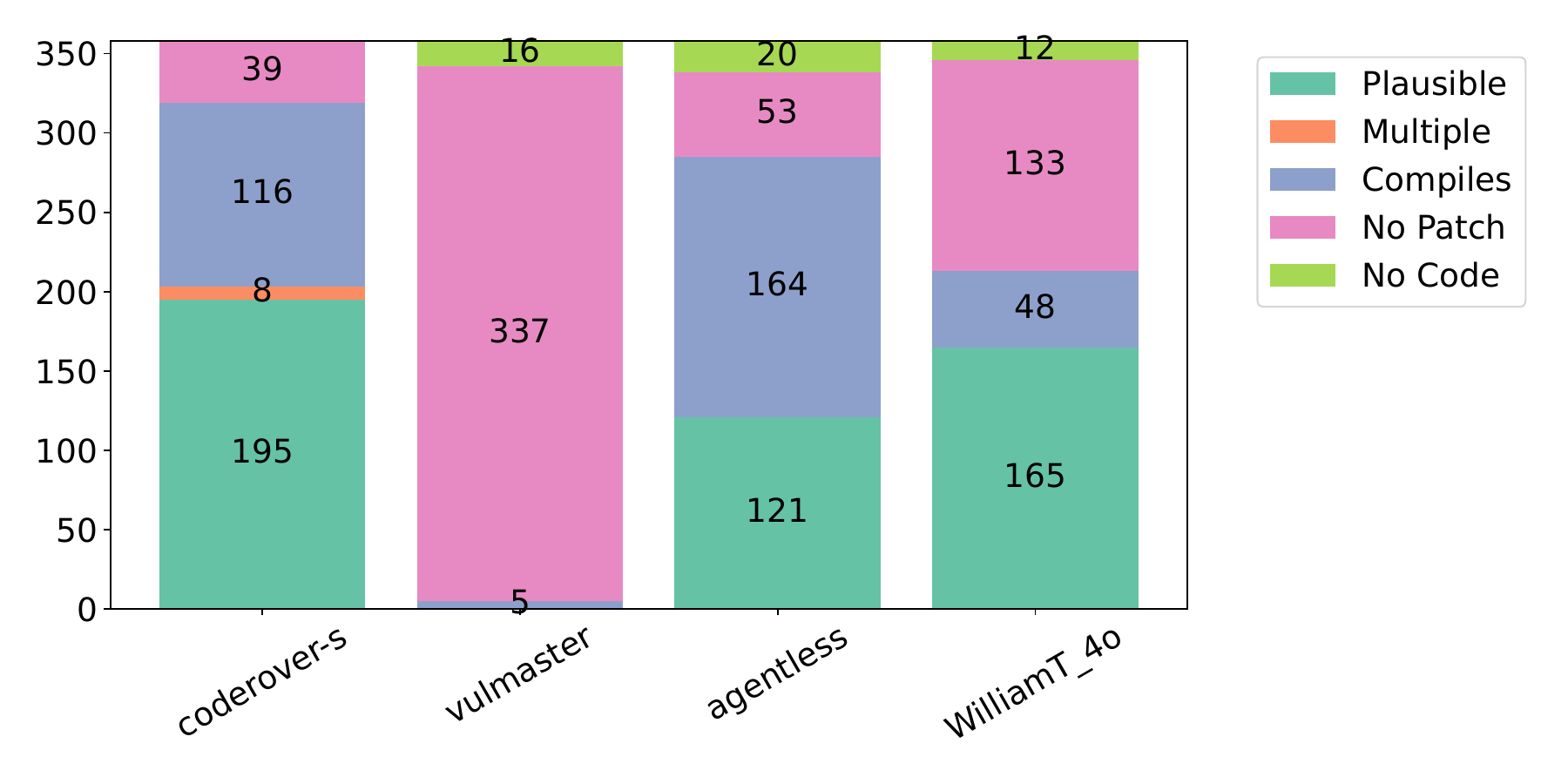}
	\caption{The fix performance of \sys (with GPT-4o) and 
    other SoTAs~\cite{zhang2024fixing,zhou2024out,xia2024automated}. 
    Multiple means the CodeRover-S takes multiple attempts to find plausible fix.}
	\label{fig:eval-rq1-perf}
\end{figure*}

We compare \sys against state-of-the-art APR tools, including 
AutoCodeRover-S~\cite{zhang2024fixing}, Agentless~\cite{xia2024automated} 
and VulMaster~\cite{zhou2024out}. As AutoCodeRover-S is not 
open-source, we import the fixes provided by AutoCodeRover-S 
for all three SoTAs for the same configuration. Additionally, 
we include \sys with same LLM model (gpt-4o-2024-08-06) 
to ensure a fair comparision.

\autoref{fig:eval-rq1-perf} presents the overall performance comparison. 
CodeRover-S achieves the highest plausible fixing rate at 54.5\%, 
followed by \sys, which successfully fixes 46.1\% of 
bugs. In contrast, Agentless performs significantly worse, with a 
fixing rate of 33.8\%, and VulMaster resolves only 5 bugs in total.
Although \sys achieves slightly lower fixing performance than 
CodeRover-S, it requires a magnitude lower resources. 
CodeRover-S attempts one patch per trial and 
conducts up to three trials per bug~\cite{zhang2024fixing}, 
which, in total, requires up to \textbf{18} attempts.
In contrast, \sys performs a single trial and applies only \textbf{one} 
patch throughout the entire fixing pipeline. This one-shot 
design significantly reduces computational and token costs 
compared to the SoTAs.

\begin{figure*}[ht]
    \centering
    \begin{subfigure}{0.6\textwidth}
        \centering
        \begin{subfigure}{0.5\textwidth}
            \centering
            \includegraphics[width=0.95\textwidth]{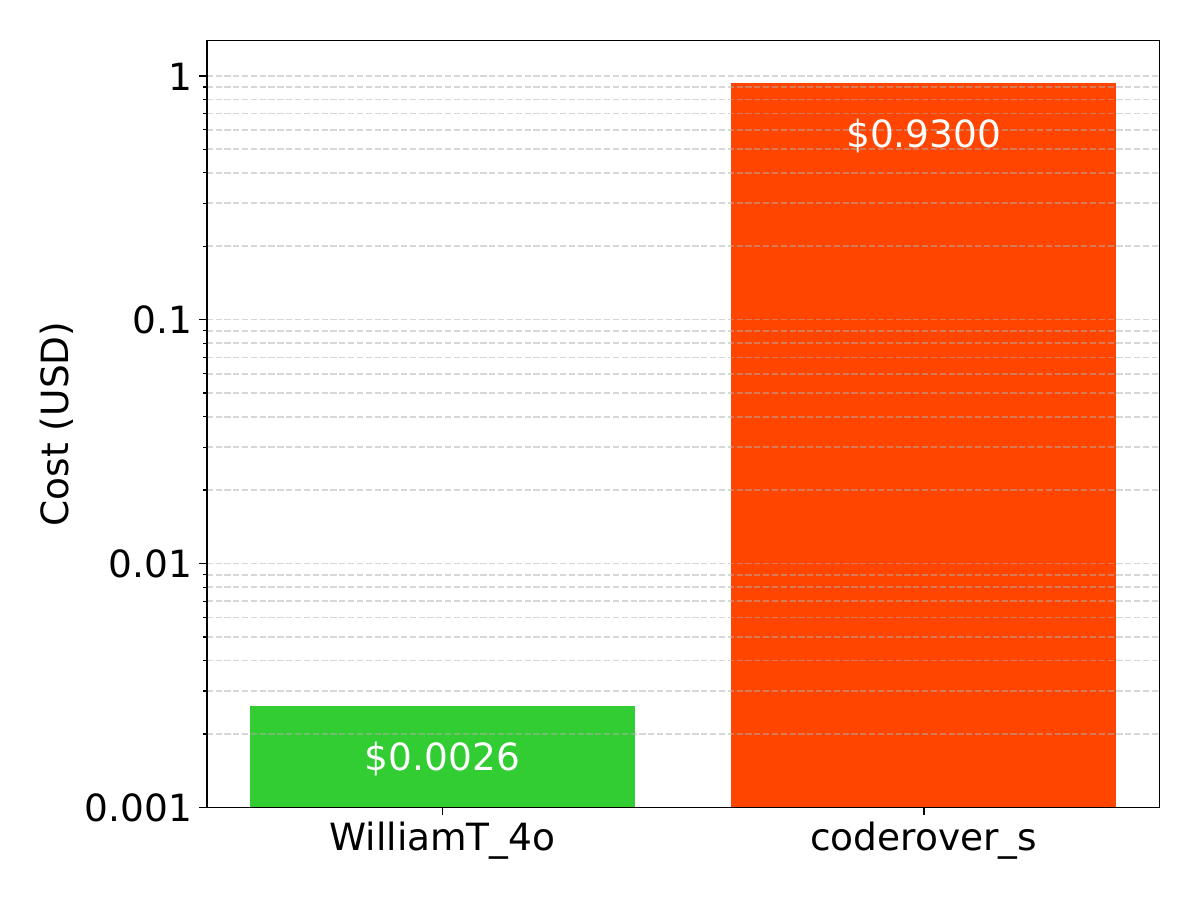}
            \caption{Average Price Cost between \sys and CodeRover-S.}
            \label{fig:eval-rq1-price-cost}
        \end{subfigure}%
        \hfill
        \begin{subfigure}{0.5\textwidth}
            \centering
            \includegraphics[width=0.95\textwidth]{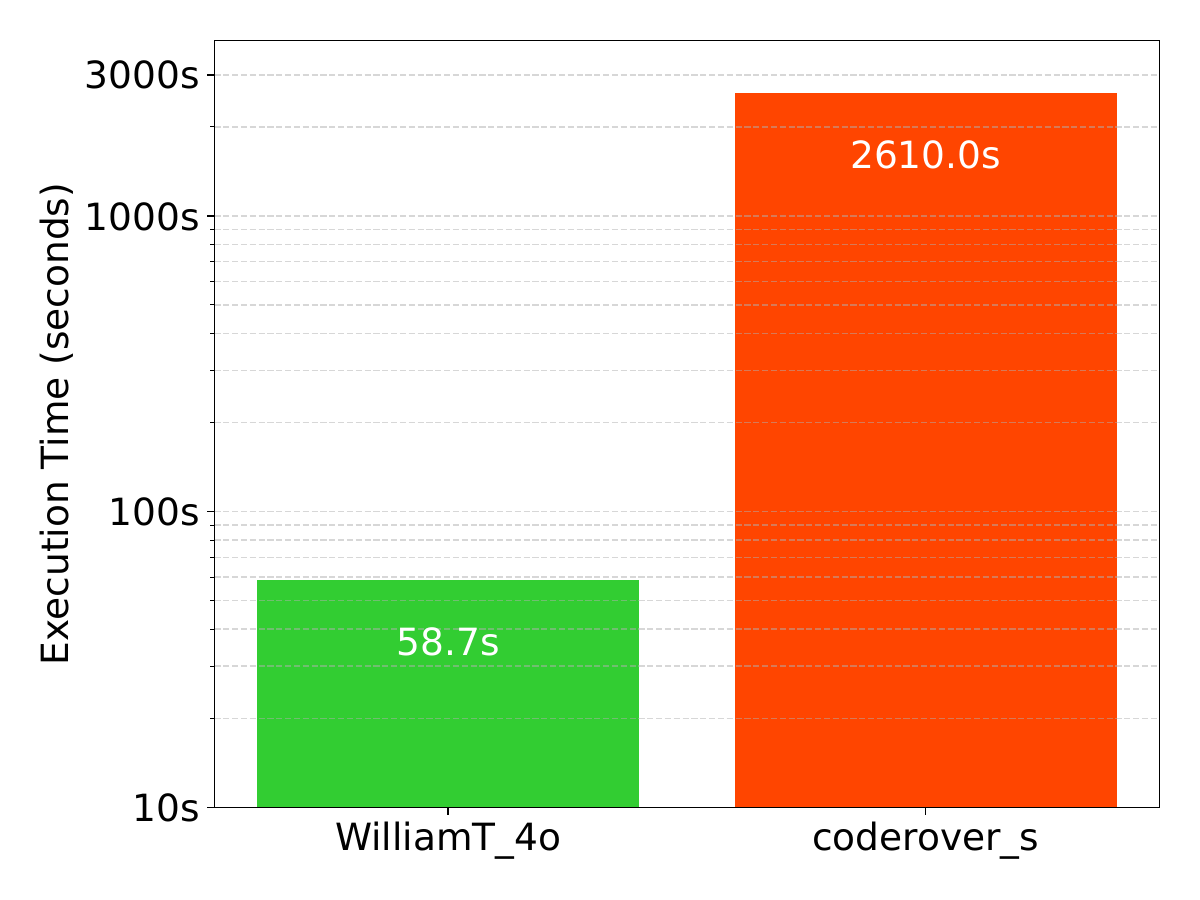}
            \caption{Average Time Cost between \sys and CodeRover-S.}
            \label{fig:eval-rq1-time-cost}
        \end{subfigure}
    \end{subfigure}%
    \hfill
    \begin{subfigure}{0.4\textwidth}
        \centering
        \includegraphics[width=0.7\textwidth,scale=0.5]{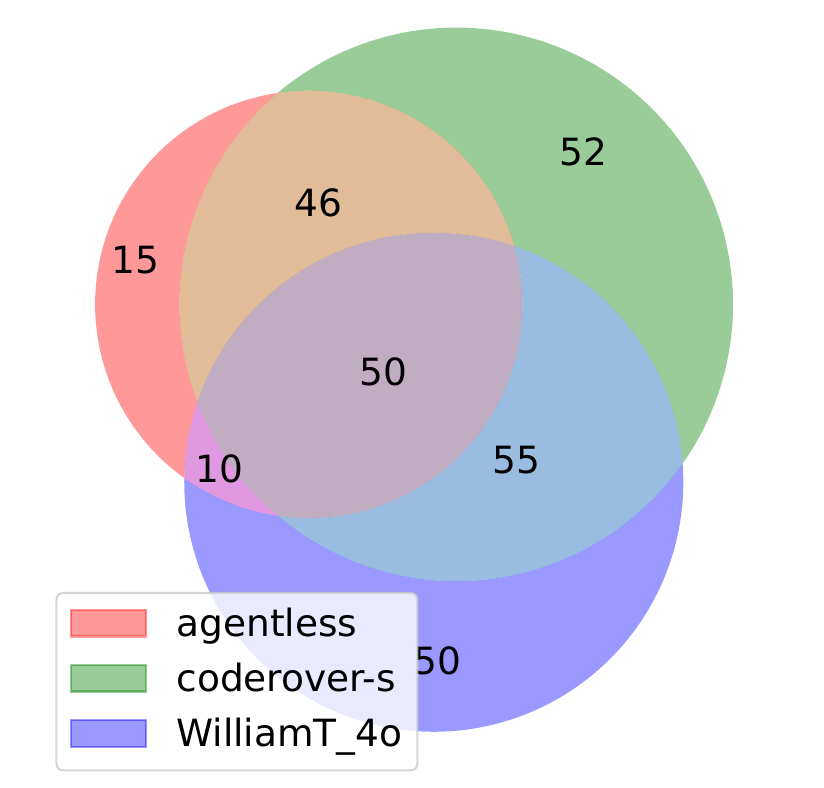}
        \caption{Venn Diagram of Plausible Patches. 
        VulMaster failed to generate any plausible patches.
        95 bugs are not fixed by any tools.}
        \label{fig:eval-rq2-venn}
    \end{subfigure}
    \label{fig:eval-overall}
\end{figure*}

\autoref{fig:eval-rq1-price-cost} compares the token and time costs 
between \sys and CodeRover-S. Both systems use the GPT-4o model 
(as of 2024-08-06). However, \sys incurs an average cost of only 
0.0026\$ per bug, whereas CodeRover-S requires approximately 0.93\$. 
In other words, \sys is able to fix \textbf{357 times} more bugs per 
dollar spent, given the same backend model.

This efficiency directly contributes to the significant disparity in 
repair time between the two systems. As shown 
in~\autoref{fig:eval-rq1-time-cost}, \sys achieves substantial time 
savings compared to CodeRover-S. On average, \sys completes the entire 
process—including preprocessing and LLM-based analysis—in under one 
minute. In contrast, CodeRover-S requires approximately 43.5 minutes 
to complete one task.
The key distinction lies in their respective repair strategies. 
CodeRover-S employs a multi-iteration repair loop, which not only 
prolongs compilation time but also imposes a considerable load on 
system resources. In contrast, \sys leverages a single-shot patching 
approach that avoids the overhead of repeated recompilation. 
This design choice significantly reduces CPU consumption and expedites 
the overall repair workflow.

\begin{tcolorbox}[colback=gray!20, colframe=black, arc=3mm, boxrule=1.5pt]
    \textbf{RQ1:} \sys saves 97.7\% CPU usage and 99.7\% Token cost, 
    while preserving over 86.7\% performance.
\end{tcolorbox}

\subsection{RQ2: What kind of bugs does \sys fix?}
\label{ssec:eval-rq2}

We begin by analyzing the overlap of plausible fixes among 
SoTA systems, as shown in~\autoref{fig:eval-rq2-venn}. Although 
CodeRover-S fixes slightly more bugs overall than \sys, the sets of 
bugs each system addresses are largely disjoint. In other words, 
\sys and CodeRover-S tend to fix different bugs. Specifically, 
\sys successfully resolves 50 bugs that neither CodeRover-S nor 
Agentless addresses, while CodeRover-S exclusively fixes 52 bugs. 
Agentless contributes 15 unique plausible fixes not covered by 
either \sys or CodeRover-S.

This low overlap highlights the complementary nature of \sys, 
suggesting it can be effectively used in combination with other 
agents. When using GPT-4o as its backend, \sys can fix all 358 bugs 
for under 0.68\$—less than the cost of fixing a single bug with 
CodeRover-S. 

Moreover, an optimized pipeline—where executing \sys is first, 
and applying CodeRover-S to cases that \sys fails—achieves 
60 additional plausible fixes, \ie 
29.6\% improvement in fixing rate compared to CodeRover-S standalone, 
while reducing the total cost by 45.9\%.

\begin{tcolorbox}[colback=gray!20, colframe=black, arc=3mm, boxrule=1.5pt]
    \textbf{RQ2:} Combining \sys and SoTA saves 45.9\% cost 
    while fixing 29.6\% more bugs.
\end{tcolorbox}

\subsection{RQ3: Does \sys scale on other LLMs?}
\label{ssec:eval-rq3}

To prove \sys's generality, we try \sys on different set of 
LLMs, including DeepSeek (R1, V3), ChatGPT (4o, o3-mini), 
Claude (3.5-Haiku and 3.7-Sonnet), 
and local LLMs that can be deployed on an 4090 GPU or even 
Mac Mini M4, 
including Gemma3 (1B, 4B, 12B, 27B).

\begin{figure*}[ht]
	\centering
	\includegraphics[width=\linewidth]{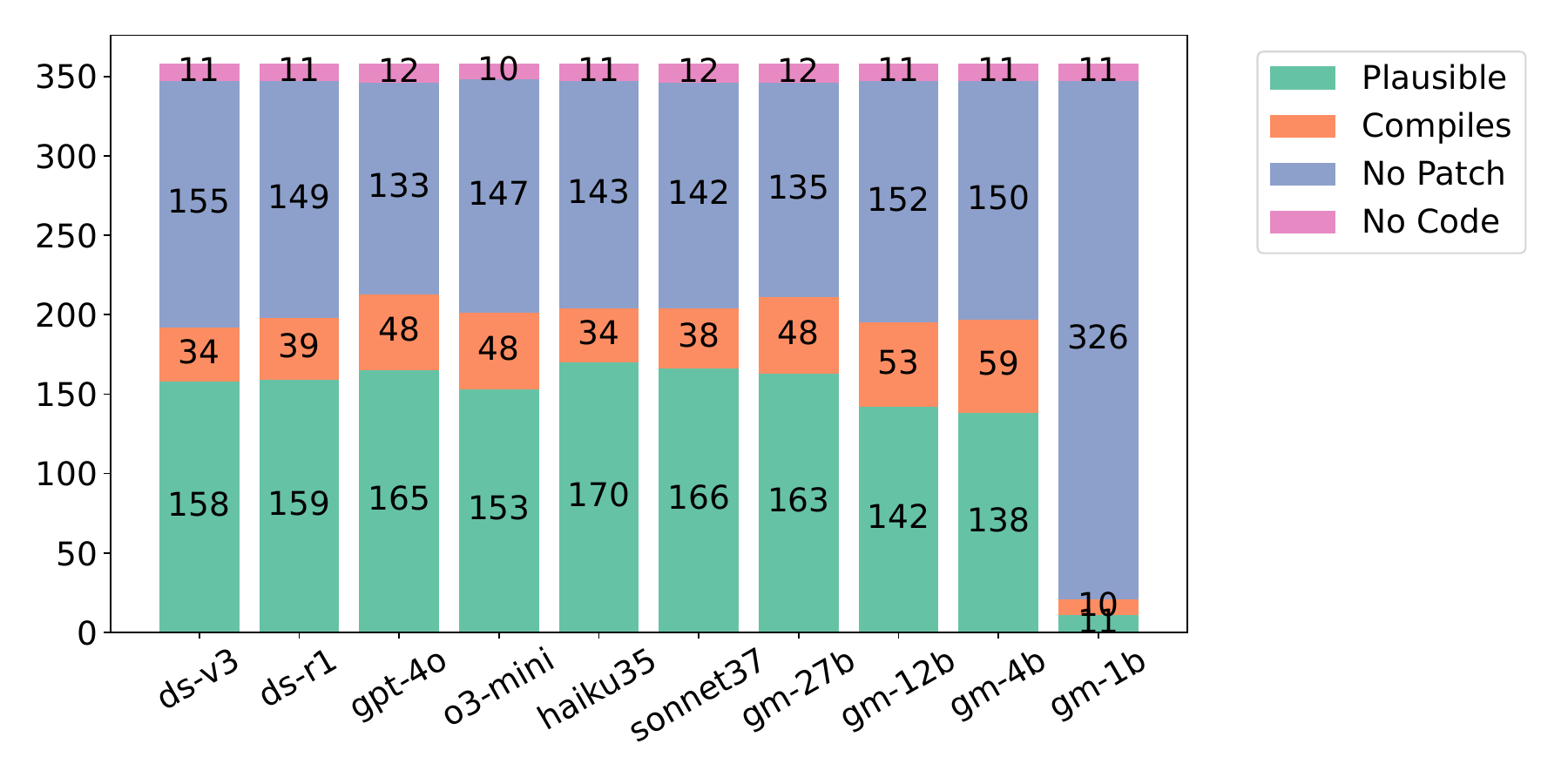}
	\caption{The \sys performance with different LLMs. 
    ds: DeepSeek, gm: Gemma3.}
	\label{fig:eval-rq3-perf}
\end{figure*}

\autoref{fig:eval-rq3-perf} shows the performance of different 
LLMs. Among all models, Claude35-haiku achieves the highest fixing 
rate at 47.5\%. Claude37-sonnet and GPT-4o follow closely, with 
fixing rates of 46.4\% and 46.1\%, respectively. 
DeepSeek also performs well, with 159 successful fixes using the 
reasoner and 158 using the chat. GPT-o3-mini performs slightly 
lower, fixing 153 bugs and achieving a success rate of 43.9\%.

Focusing on smaller language models, particularly the Gemma3 
series, we observe that they perform comparably well relative to 
frontier models. Notably, Gemma3:27B ranks second overall, achieving 
96.4\% of GPT-4o’s performance. While the 12B and 4B versions of 
Gemma3 exhibit a noticeable drop in performance, they still manage 
to fix 39.4\% and 38.0\% of bugs, respectively. 

Importantly, the Gemma3:27B model is lightweight enough to run on a 
single consumer-grade GPU, such as an RTX 4090, or even on a 
Mac Mini M4. This demonstrates not only the scalability of \sys 
but also its practical feasibility for deployment on local developer 
machines.

\begin{figure*}[h!]
	\centering
	\begin{subfigure}{0.5\textwidth}
			\centering
			\includegraphics[width=0.95\textwidth]{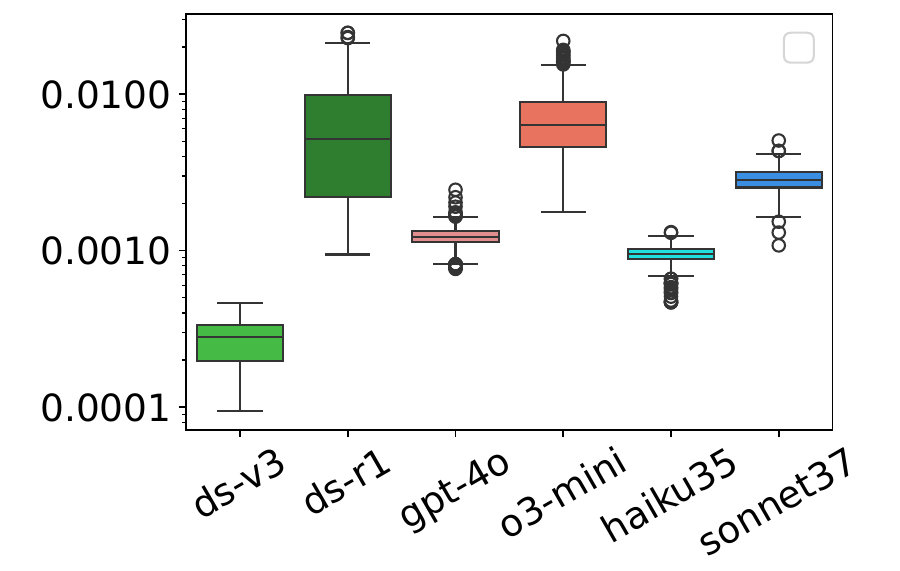}
			\caption{Price Estimation (USD)}
	\end{subfigure}%
	\begin{subfigure}{0.5\textwidth}
			\centering
			\includegraphics[width=0.95\textwidth]{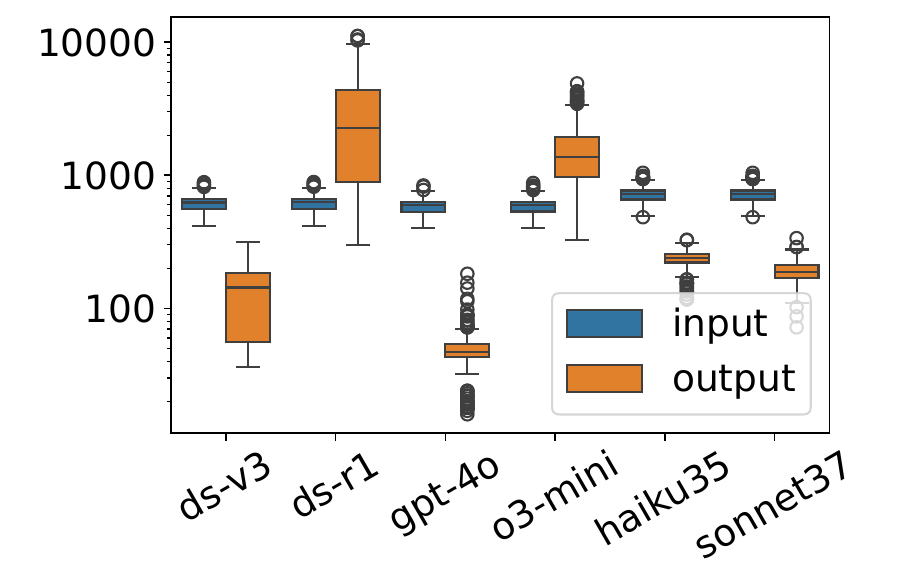}
			\caption{Token Consumption}
	\end{subfigure}%
    \caption{Token and Price Cost for different LLMs. 
    ds: DeepSeek, gm: Gemma3.}
    \label{fig:eval-rq3-cost}
\end{figure*}

We further analyze the cost efficiency of various frontier LLM backends. 
All cost are measured in cent USD, \ie 0.01 USD.
As shown in \autoref{fig:eval-rq3-cost}, DeepSeek-V3 achieves the 
highest efficiency, with a cost of less than 0.03 cent per bug. 
Claude35-Haiku, while attaining the best fixing performance, 
maintains a moderate cost of approximately 0.09 cent per bug. 
In contrast, reasoning models such as DeepSeek-R1 and GPT-o3-mini 
exhibit significantly higher costs, at 0.70 cent and 0.72 cent per bug, 
respectively—5.7× and 5.9× more expensive than Claude35-Haiku—without 
providing notable improvements over non-reasoning LLMs.

The primary reason for the increased cost is the excessive output token. 
Although all models consume a similar number of input 
tokens, their output token consumption varies significantly. 
DeepSeek-V3 and GPT-4o generate an average of 
131 and 49 output tokens per bug, respectively, whereas Claude35-Haiku 
averages 236 output tokens.
Among reasoning models, DeepSeek-R1 and GPT-o3-mini produce 
substantially more output, averaging 3,060 and 1,564 output 
tokens per bug, respectively. An exception is Claude-3-7-Sonnet, a 
hybrid reasoning model that performs on-demand reasoning. Its average 
of 191 output tokens suggests that our task setup largely avoids heavy 
reasoning.
Importantly, this large increase in output tokens does not correlate 
with better fixing performance. In fact, Claude35-Haiku, a non-reasoning 
model, achieves the best results across all models.

These findings highlight the potential of \sys as an agent: by 
providing a strong template, it can eliminate the need for expensive 
reasoning, enabling non-reasoning models to outperform their 
reasoning-based counterparts.

\begin{figure*}[!h]
	\centering
	\includegraphics[width=\linewidth]{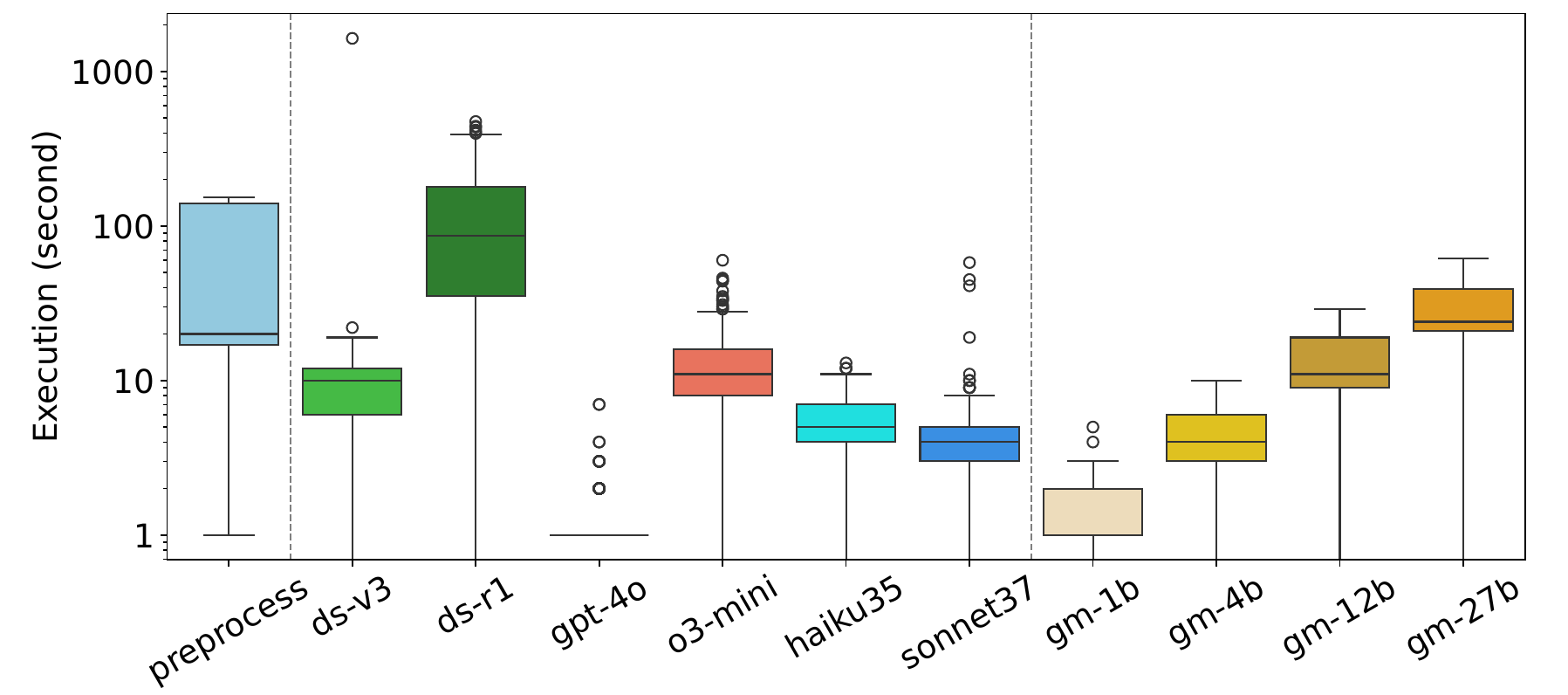}
	\caption{Execution cost for preprocessing and different LLMs to analysis variable. 
    The dot lines split the preprocessing, frontier LLMs and local LLMs.}
	\label{fig:eval-rq3-exec}
\end{figure*}

We further decompose the time cost associated with bug analysis in 
\sys. As a one-shot Automated Program Repair (APR) agent, \sys requires 
only a preprocessing phase to extract necessary contextual 
information—independent of the choice of LLM backend—and a subsequent 
LLM-based key variable analysis. The breakdown is illustrated in 
\autoref{fig:eval-rq3-exec}.

On average, the preprocessing takes less than 1 minute, 
primarily due to Python dependency compilation and source code analysis. 
We also evaluate the execution time of state-of-the-art LLMs used for 
the key variable analysis. Non-reasoning models typically complete 
this step in under 10 seconds, whereas reasoning-capable models range 
from 13 seconds (\ie o3-mini) to 120 seconds (\ie DeepSeek-R1). 
For local execution, we benchmarked the Gemma-3 variants on an RTX 4090 
GPU, observing runtimes between 1 and 13 seconds.
Overall, the majority of \sys’s execution time is attributed to 
preprocessing. Even when using the slowest reasoning model, \sys 
consistently completes the bug-fixing task in under 3 
minutes—illustrating a substantial improvement over CodeRover-S, 
which requires approximately 43.5 minutes.

\begin{tcolorbox}[colback=gray!20, colframe=black, arc=3mm, boxrule=1.5pt]
    \textbf{RQ3:} 
    \sys performs well with frontier models while remaining 
    scalable to smaller, more cost-effective LLMs.
\end{tcolorbox}

\section{Discussion}
\label{sec:discussion}

\sys produces a correct patch when both the crash site analysis 
accurately identifies relevant variables and the patch insertion is 
semantically valid within the target code. In this section, we examine 
potential sources of false positives in \sys, focusing on internal 
and external threats to patch plausibility. Internally, 
inaccuracies may arise from \emph{incorrect crash site analysis} or 
\emph{semantically disruptive patch insertion}. 
Externally, the \emph{imprecise plausible fix metric} used during 
evaluation may fail to distinguish correct from incorrect patches.

\noindent \textbf{Incorrect Crash Site Analysis.}
While \sys demonstrates a promising plausible fix rate, we observe 
cases where key variables returned by the LLM are inaccurate. 
This issue can stem from two primary factors. First, the code 
snippet provided to the LLM may lack the correct variable due 
to limited context, especially when the required variables lies 
beyond the final crash frame. Second, the precision of variable identification 
is constrained by the LLM’s capabilities and the effectiveness of 
prompt engineering. Addressing these limitations—potentially 
through enhanced context retrieval or improved prompt 
strategies—is left as future work. 
We further discuss its impact in~\autoref{sec:append-plausible-fix}.

\noindent \textbf{Semantically Disruptive Patch Insertion.}
\sys currently inserts the generated patch one line above the crash 
site. While this approach may work in principle, it can lead to 
semantic errors and compilation failures in practice. For instance, 
if the crash occurs within a conditional block (\eg inside an 
\lstinline{if (...)} statement), naive insertion can break control 
flow and invalidate the program structure. 
In our evaluation, 'No Patch' failures account for approximately 
37\% of all bugs, with many caused by such semantic violations.
Incorporating LLM-guided semantic-aware insertion strategies may 
offer a viable solution to mitigate this issue.

\noindent \textbf{Imprecise Plausible Metric.} 
Our evaluation follows the standard plausibility criterion from 
prior work~\cite{zhang2024fixing}, which verifies that a generated 
patch prevents the original crash. However, this metric does not 
ensure that the program's functionality is preserved. 
Consequently, it may overestimate the number of truly correct fixes 
and introduce false positives across all evaluated systems. This 
limitation highlights the need for more rigorous metrics that account 
for semantic correctness and broader behavioral preservation.

\section{Conclusion}
\label{sec:conclusion}

The increasing number of discovered security bugs creates a strong 
need for better automated program repair (APR). Instead of relying 
on complex root cause analysis like most existing APR tools, we 
propose \emph{crash-site repair} to simplify the repair process 
and improve precision. We also introduce a \emph{template-guided patch 
generation} technique to reduce the high cost of using LLMs, 
while preserving the repairs effectiveness.
We implement our prototype \sys, and evaluate it against 
state-of-the-art tools. Our results show that combining \sys with the 
best-performing tool, CodeRover-S, reduces token usage by 29.6\% 
and improves the fixing rate by 45.9\% compared to CodeRover-S alone. 
In addition, \sys works well with local LLMs on a Mac Mini, showing 
its potential for low-cost, local development.
We promise to fully open-source \sys upon paper acceptance.




\bibliographystyle{plain}

\bibliography{paper}

\begin{thebibliography}{10}

\bibitem{grok}
X~AI.
\newblock Grok.
\newblock \url{https://x.ai/}, 2025.

\bibitem{bhandari2021cvefixes}
Guru Bhandari, Amara Naseer, and Leon Moonen.
\newblock Cvefixes: automated collection of vulnerabilities and their fixes
  from open-source software.
\newblock In {\em Proceedings of the 17th International Conference on
  Predictive Models and Data Analytics in Software Engineering}, pages 30--39,
  2021.

\bibitem{chrome_exploit_bh_24_2}
Liu Bohan and Shi Haibin.
\newblock Escape modern web-based app sandbox from site-isolation perspective.
\newblock
  \url{https://i.blackhat.com/Asia-24/Presentations/Asia-24-Liu-The-Hole-in-Sandbox.pdf},
  2024.

\bibitem{chrome_exploit_bh_23_1}
Liu Bohan and Wang Zheng.
\newblock Reviving jit vulnerabilities: Unleashing the power of maglev compiler
  bugs on chrome browser.
\newblock
  \url{https://i.blackhat.com/EU-23/Presentations/EU-23-Liu-Reviving-JIT-Vulnerabilities.pdf},
  2023.

\bibitem{bulekov2024hyperpill}
Alexander Bulekov, Qiang Liu, Manuel Egele, and Mathias Payer.
\newblock $\{$HYPERPILL$\}$: Fuzzing for hypervisor-bugs by leveraging the
  hardware virtualization interface.
\newblock In {\em 33rd USENIX Security Symposium (USENIX Security 24)}, pages
  919--935, 2024.

\bibitem{oss_fuzz_blog}
Oliver Chang and OSS-Fuzz team.
\newblock Taking the next step: Oss-fuzz in 2023, 2023.

\bibitem{chen2019sequencer}
Zimin Chen, Steve Kommrusch, Michele Tufano, Louis-No{\"e}l Pouchet, Denys
  Poshyvanyk, and Martin Monperrus.
\newblock Sequencer: Sequence-to-sequence learning for end-to-end program
  repair.
\newblock {\em IEEE Transactions on Software Engineering}, 47(9):1943--1959,
  2019.

\bibitem{chrome_checklist}
chromium.
\newblock Top security things for chromies to remember.
\newblock
  \url{https://chromium.googlesource.com/chromium/src/+/lkgr/docs/security/checklist.md},
  2025.

\bibitem{cgc_binary}
DARPA.
\newblock Cyber grand challenge - datasets, 2025.

\bibitem{fan2020ac}
Jiahao Fan, Yi~Li, Shaohua Wang, and Tien~N Nguyen.
\newblock Ac/c++ code vulnerability dataset with code changes and cve
  summaries.
\newblock In {\em Proceedings of the 17th international conference on mining
  software repositories}, pages 508--512, 2020.

\bibitem{feng2020codebert}
Zhangyin Feng, Daya Guo, Duyu Tang, Nan Duan, Xiaocheng Feng, Ming Gong, Linjun
  Shou, Bing Qin, Ting Liu, Daxin Jiang, et~al.
\newblock Codebert: A pre-trained model for programming and natural languages.
\newblock {\em arXiv preprint arXiv:2002.08155}, 2020.

\bibitem{fioraldi2020afl++}
Andrea Fioraldi, Dominik Maier, Heiko Ei{\ss}feldt, and Marc Heuse.
\newblock Afl++ combining incremental steps of fuzzing research.
\newblock In {\em Proceedings of the 14th USENIX Conference on Offensive
  Technologies}, pages 10--10, 2020.

\bibitem{fleischer2023actor}
Marius Fleischer, Dipanjan Das, Priyanka Bose, Weiheng Bai, Kangjie Lu, Mathias
  Payer, Christopher Kruegel, and Giovanni Vigna.
\newblock $\{$ACTOR$\}$:$\{$Action-Guided$\}$ kernel fuzzing.
\newblock In {\em 32nd USENIX Security Symposium (USENIX Security 23)}, pages
  5003--5020, 2023.

\bibitem{gao2021beyond}
Xiang Gao, Bo~Wang, Gregory~J Duck, Ruyi Ji, Yingfei Xiong, and Abhik
  Roychoudhury.
\newblock Beyond tests: Program vulnerability repair via crash constraint
  extraction.
\newblock {\em ACM Transactions on Software Engineering and Methodology
  (TOSEM)}, 30(2):1--27, 2021.

\bibitem{clusterfuzz}
google.
\newblock Clusterfuzz.
\newblock \url{https://google.github.io/clusterfuzz/}, 2023.

\bibitem{oss_fuzz}
Google.
\newblock Oss-fuzz - continuous fuzzing for open source software, 2025.

\bibitem{syzkaller}
Google.
\newblock syzkaller is an unsupervised coverage-guided kernel fuzzer.
\newblock \url{https://github.com/google/syzkaller}, 2025.

\bibitem{gemma3}
Google.
\newblock Welcome gemma 3: Google's all new multimodal, multilingual, long
  context open llm.
\newblock \url{http://huggingface.co/blog/gemma3}, 2025.

\bibitem{gross2023fuzzilli}
Samuel Gro{\ss}, Simon Koch, Lukas Bernhard, Thorsten Holz, and Martin Johns.
\newblock Fuzzilli: Fuzzing for javascript jit compiler vulnerabilities.
\newblock In {\em NDSS}, 2023.

\bibitem{guo2025deepseek}
Daya Guo, Dejian Yang, Haowei Zhang, Junxiao Song, Ruoyu Zhang, Runxin Xu,
  Qihao Zhu, Shirong Ma, Peiyi Wang, Xiao Bi, et~al.
\newblock Deepseek-r1: Incentivizing reasoning capability in llms via
  reinforcement learning.
\newblock {\em arXiv preprint arXiv:2501.12948}, 2025.

\bibitem{huang2023empirical}
Kai Huang, Xiangxin Meng, Jian Zhang, Yang Liu, Wenjie Wang, Shuhao Li, and
  Yuqing Zhang.
\newblock An empirical study on fine-tuning large language models of code for
  automated program repair.
\newblock In {\em 2023 38th IEEE/ACM International Conference on Automated
  Software Engineering (ASE)}, pages 1162--1174. IEEE, 2023.

\bibitem{huang2025template}
Kai Huang, Jian Zhang, Xiangxin Meng, and Yang Liu.
\newblock Template-guided program repair in the era of large language models.
\newblock ICSE, 2025.

\bibitem{chrome_exploit_bh_22_1}
Rong Jian and Guang Gong.
\newblock Another way to talk with browser : Exploiting chrome at network
  layer.
\newblock
  \url{https://i.blackhat.com/USA-22/Thursday/US-22-Rong-Another_Way_to_Talk_with_Browser_Exploiting_Chrome_at_Network_Layer.pdf},
  2022.

\bibitem{too_many_bugs_fix}
Angelos Keromytis.
\newblock Recommendations from the workshop on open-source software security
  initiative.
\newblock
  \url{https://bpb-us-e1.wpmucdn.com/sites.gatech.edu/dist/a/2878/files/2022/10/OSSI-Final-Report.pdf},
  2022.

\bibitem{libfuzzer}
libfuzzer.
\newblock libfuzzer.
\newblock \url{https://llvm.org/docs/LibFuzzer.html}, 2023.

\bibitem{liu2024deepseek}
Aixin Liu, Bei Feng, Bing Xue, Bingxuan Wang, Bochao Wu, Chengda Lu, Chenggang
  Zhao, Chengqi Deng, Chenyu Zhang, Chong Ruan, et~al.
\newblock Deepseek-v3 technical report.
\newblock {\em arXiv preprint arXiv:2412.19437}, 2024.

\bibitem{liu2023videzzo}
Qiang Liu, Flavio Toffalini, Yajin Zhou, and Mathias Payer.
\newblock Videzzo: Dependency-aware virtual device fuzzing.
\newblock In {\em 2023 IEEE Symposium on security and privacy (SP)}, pages
  3228--3245. IEEE, 2023.

\bibitem{matruman}
Zheyu Ma, Qiang Liu, Zheming Li, Tingting Yin, Wende Tan, Chao Zhang, and
  Mathias Payer.
\newblock Truman: Constructing device behavior models from os drivers to fuzz
  virtual devices.

\bibitem{kernel_exploit_bh_22}
Bogaard Martijn and Geist Dana.
\newblock Achieving linux kernel code execution through a malicious usb device.
\newblock
  \url{https://i.blackhat.com/EU-21/Thursday/EU-21-Bogaard_Geist_Achieving_Linux_Kernel_Code_Execution_Through_A_Malicious_USB_Device.pdf},
  2021.

\bibitem{mei2024arvo}
Xiang Mei, Pulkit~Singh Singaria, Jordi Del~Castillo, Haoran Xi, Tiffany Bao,
  Ruoyu Wang, Yan Shoshitaishvili, Adam Doup{\'e}, Hammond Pearce, Brendan
  Dolan-Gavitt, et~al.
\newblock Arvo: Atlas of reproducible vulnerabilities for open source software.
\newblock {\em arXiv preprint arXiv:2408.02153}, 2024.

\bibitem{llama4}
Meta.
\newblock The llama 4 herd: The beginning of a new era of natively multimodal
  ai innovation.
\newblock \url{https://ai.meta.com/blog/llama-4-multimodal-intelligence/},
  2025.

\bibitem{chrome_exploit_bh_24_1}
Wang Nan and Xiao Zhenghang.
\newblock Exploit chrome and firefox four times.
\newblock
  \url{https://i.blackhat.com/BH-US-24/Presentations/US24-Xiao-Super-Hat-Trick-Exploit-Chrome-and-Firefox.pdf},
  2024.

\bibitem{hack_stack1996}
Aleph One.
\newblock Hacking the stack for fun and profit.
\newblock {\em Phrack Magazine}, 1996.

\bibitem{chatgpt}
OpenAI.
\newblock Chatgpt.
\newblock \url{https://chatgpt.com}, 2025.

\bibitem{pan2021v}
Gaoning Pan, Xingwei Lin, Xuhong Zhang, Yongkang Jia, Shouling Ji, Chunming Wu,
  Xinlei Ying, Jiashui Wang, and Yanjun Wu.
\newblock V-shuttle: Scalable and semantics-aware hypervisor virtual device
  fuzzing.
\newblock In {\em Proceedings of the 2021 ACM SIGSAC Conference on Computer and
  Communications Security}, pages 2197--2213, 2021.

\bibitem{Payer18SS3P}
Mathias Payer.
\newblock {\em {Software Security: Principles, Policies, and Protection}}.
\newblock HexHive Books, 0.37 edition, July 2021.

\bibitem{asan_callstack}
Alexander Potapenko.
\newblock Addresssanitizercallstack, 2015.

\bibitem{chrome_third_party_bug}
External reporter.
\newblock Security: heap-buffer-overflow in libavif when decode the crafted
  avif file, 2024.

\bibitem{apple_third_party_bug}
Apple Security.
\newblock ios 16.1.1 and ipados 16.1.1, 2024.

\bibitem{serebryany2012addresssanitizer}
Konstantin Serebryany, Derek Bruening, Alexander Potapenko, and Dmitriy Vyukov.
\newblock $\{$AddressSanitizer$\}$: A fast address sanity checker.
\newblock In {\em 2012 USENIX annual technical conference (USENIX ATC 12)},
  pages 309--318, 2012.

\bibitem{chrome_exploit_bh_22_2}
Röttger Stephen.
\newblock Breaking the chrome sandbox with mojo.
\newblock
  \url{https://i.blackhat.com/USA-22/Wednesday/US-22-Roettger_Breaking_the_Chrome_Sandbox_with_Mojo.pdf},
  2022.

\bibitem{memory_safety_android}
Jeffrey~Vander Stoep.
\newblock Memory safe languages in android 13, 2022.

\bibitem{memory_corruption_exploitable}
Maddie Stone.
\newblock The more you know, the more you know you don’t know - a year in
  review of 0-days used in-the-wild in 2021, 2022.

\bibitem{syzbot_open_bugs}
Syzbot.
\newblock Syzbot open bugs, 2025.

\bibitem{tang2024code}
Hao Tang, Keya Hu, Jin Zhou, Si~Cheng Zhong, Wei-Long Zheng, Xujie Si, and
  Kevin Ellis.
\newblock Code repair with llms gives an exploration-exploitation tradeoff.
\newblock {\em Advances in Neural Information Processing Systems},
  37:117954--117996, 2024.

\bibitem{memory_safety_chrome}
Chromium~Security Team.
\newblock Memory safety in chromium, 2025.

\bibitem{team2024gemma}
Gemma Team, Thomas Mesnard, Cassidy Hardin, Robert Dadashi, Surya Bhupatiraju,
  Shreya Pathak, Laurent Sifre, Morgane Rivi{\`e}re, Mihir~Sanjay Kale,
  Juliette Love, et~al.
\newblock Gemma: Open models based on gemini research and technology.
\newblock {\em arXiv preprint arXiv:2403.08295}, 2024.

\bibitem{qwq32b}
Qwen Team.
\newblock Qwq-32b: Embracing the power of reinforcement learning, March 2025.

\bibitem{tufano2019empirical}
Michele Tufano, Cody Watson, Gabriele Bavota, Massimiliano~Di Penta, Martin
  White, and Denys Poshyvanyk.
\newblock An empirical study on learning bug-fixing patches in the wild via
  neural machine translation.
\newblock {\em ACM Transactions on Software Engineering and Methodology
  (TOSEM)}, 28(4):1--29, 2019.

\bibitem{wachterdumpling}
Liam Wachter, Julian Gremminger, Christian Wressnegger, Mathias Payer, and
  Flavio Toffalini.
\newblock Dumpling: Fine-grained differential javascript engine fuzzing.

\bibitem{wang2024syztrust}
Qinying Wang, Boyu Chang, Shouling Ji, Yuan Tian, Xuhong Zhang, Binbin Zhao,
  Gaoning Pan, Chenyang Lyu, Mathias Payer, Wenhai Wang, et~al.
\newblock Syztrust: State-aware fuzzing on trusted os designed for iot devices.
\newblock In {\em 2024 IEEE Symposium on Security and Privacy (SP)}, pages
  2310--2387. IEEE, 2024.

\bibitem{macos_exploit_bh_21}
Reguła Wojciech and Fitzl Csaba.
\newblock 20+ ways to bypass your macos privacy mechanisms.
\newblock
  \url{https://i.blackhat.com/USA21/Wednesday-Handouts/US-21-Regula-20-Plus-Ways-to-Bypass-Your-macOS-Privacy-Mechanisms.pdf},
  2021.

\bibitem{xia2022less}
Chunqiu~Steven Xia and Lingming Zhang.
\newblock Less training, more repairing please: revisiting automated program
  repair via zero-shot learning.
\newblock In {\em Proceedings of the 30th ACM Joint European Software
  Engineering Conference and Symposium on the Foundations of Software
  Engineering}, pages 959--971, 2022.

\bibitem{xia2024automated}
Chunqiu~Steven Xia and Lingming Zhang.
\newblock Automated program repair via conversation: Fixing 162 out of 337 bugs
  for \$0.42 each using chatgpt.
\newblock In {\em Proceedings of the 33rd ACM SIGSOFT International Symposium
  on Software Testing and Analysis}, pages 819--831, 2024.

\bibitem{xu2020freedom}
Wen Xu, Soyeon Park, and Taesoo Kim.
\newblock Freedom: Engineering a state-of-the-art dom fuzzer.
\newblock In {\em Proceedings of the 2020 ACM SIGSAC Conference on Computer and
  Communications Security}, pages 971--986, 2020.

\bibitem{qwen2.5}
An~Yang, Baosong Yang, Beichen Zhang, Binyuan Hui, Bo~Zheng, Bowen Yu,
  Chengyuan Li, Dayiheng Liu, Fei Huang, Haoran Wei, Huan Lin, Jian Yang,
  Jianhong Tu, Jianwei Zhang, Jianxin Yang, Jiaxi Yang, Jingren Zhou, Junyang
  Lin, Kai Dang, Keming Lu, Keqin Bao, Kexin Yang, Le~Yu, Mei Li, Mingfeng Xue,
  Pei Zhang, Qin Zhu, Rui Men, Runji Lin, Tianhao Li, Tianyi Tang, Tingyu Xia,
  Xingzhang Ren, Xuancheng Ren, Yang Fan, Yang Su, Yichang Zhang, Yu~Wan,
  Yuqiong Liu, Zeyu Cui, Zhenru Zhang, and Zihan Qiu.
\newblock Qwen2.5 technical report.
\newblock {\em arXiv preprint arXiv:2412.15115}, 2024.

\bibitem{yin2023kextfuzz}
Tingting Yin, Zicong Gao, Zhenghang Xiao, Zheyu Ma, Min Zheng, and Chao Zhang.
\newblock $\{$KextFuzz$\}$: Fuzzing $\{$macOS$\}$ kernel $\{$EXTensions$\}$ on
  apple silicon via exploiting mitigations.
\newblock In {\em 32nd USENIX Security Symposium (USENIX Security 23)}, pages
  5039--5054, 2023.

\bibitem{afl}
Michal Zalewski.
\newblock american fuzzy lop.
\newblock \url{https://lcamtuf.coredump.cx/afl/}, 2013.

\bibitem{zhang2024autocoderover}
Yuntong Zhang, Haifeng Ruan, Zhiyu Fan, and Abhik Roychoudhury.
\newblock Autocoderover: Autonomous program improvement.
\newblock In {\em Proceedings of the 33rd ACM SIGSOFT International Symposium
  on Software Testing and Analysis}, pages 1592--1604, 2024.

\bibitem{zhang2024fixing}
Yuntong Zhang, Jiawei Wang, Dominic Berzin, Martin Mirchev, Dongge Liu,
  Abhishek Arya, Oliver Chang, and Abhik Roychoudhury.
\newblock Fixing security vulnerabilities with ai in oss-fuzz.
\newblock {\em arXiv preprint arXiv:2411.03346}, 2024.

\bibitem{zheng2025mendelfuzz}
Han Zheng, Flavio Toffalini, Marcel B{\"o}hme, and Mathias Payer.
\newblock Mendelfuzz: The return of the deterministic stage.
\newblock 2025.

\bibitem{zheng2023fishfuzz}
Han Zheng, Jiayuan Zhang, Yuhang Huang, Zezhong Ren, He~Wang, Chunjie Cao,
  Yuqing Zhang, Flavio Toffalini, and Mathias Payer.
\newblock $\{$FISHFUZZ$\}$: Catch deeper bugs by throwing larger nets.
\newblock In {\em 32nd USENIX Security Symposium (USENIX Security 23)}, pages
  1343--1360, 2023.

\bibitem{zhou2022minerva}
Chijin Zhou, Quan Zhang, Mingzhe Wang, Lihua Guo, Jie Liang, Zhe Liu, Mathias
  Payer, and Yu~Jiang.
\newblock Minerva: browser api fuzzing with dynamic mod-ref analysis.
\newblock In {\em Proceedings of the 30th ACM Joint European Software
  Engineering Conference and Symposium on the Foundations of Software
  Engineering}, pages 1135--1147, 2022.

\bibitem{zhou2024out}
Xin Zhou, Kisub Kim, Bowen Xu, DongGyun Han, and David Lo.
\newblock Out of sight, out of mind: Better automatic vulnerability repair by
  broadening input ranges and sources.
\newblock In {\em Proceedings of the IEEE/ACM 46th International Conference on
  Software Engineering}, pages 1--13, 2024.

\bibitem{zhu2024crossfire}
Jiaxun Zhu, Minghao Lin, Tingting Yin, Zechao Cai, Yu~Wang, Rui Chang, and
  Wenbo Shen.
\newblock Crossfire: Fuzzing macos cross-xpu memory on apple silicon.
\newblock In {\em Proceedings of the 2024 on ACM SIGSAC Conference on Computer
  and Communications Security}, pages 3749--3762, 2024.

\end{thebibliography}


\appendix

\section{The Plausible Fix and Actual Fix}
\label{sec:append-plausible-fix}

In this section, we discuss the differences between a \emph{plausible fix} 
and a \emph{successful fix}.

Current practice~\cite{zhang2024fixing} uses the \emph{plausible fix} 
as the primary metric to evaluate the capability of APR 
tools. Specifically, a patch is considered plausible if the 
patched program, when run with the proof-of-concept (PoC) input, 
does not crash. However, this metric is insufficient. We summarize 
the potential issues with this 
approach in~\autoref{tab:appendix-metric-compare}.

\begin{table}[h!]
    \resizebox{\textwidth}{!} {
    \begin{tabular}{l|rrr|r}
    \toprule
    \multirow{2}{*}{Metric} & \multicolumn{3}{c|}{Taking PoC as Input}             & \multirow{2}{*}{Other Input} \\
                            & Not Exploitable & No Early Exit & Same Results &                              \\ \midrule
    Plausible               & \cmark                & \xmark               & \xmark$^{*}$         & \xmark                          \\
    Same Execution          & \cmark                & \cmark               & \cmark           & \xmark                          \\
    Manual Review           & \cmark                & \cmark                & \cmark           & \cmark                           \\ \midrule
    \end{tabular}
    }
    \caption{Comparision between different metrics. *: \sys guarantees that 
    any plausible patch yeild same results if not early exit.}
    \label{tab:appendix-metric-compare}
\end{table}

In particular, the current \emph{plausible fix} metric only verifies 
that, for the given PoC input, the program is no longer exploitable. 
On the one hand, the program may still exhibit issues such as \emph{early exit} 
or \emph{different execution results} when given the same PoC. 
On the other hand, other inputs may still bypass the patch and trigger 
the original vulnerability.
This situation highlights the need for a more robust metric—one that 
ensures the program, whether executed with the PoC or with other 
inputs, is \emph{no longer exploitable}, avoids \emph{early exits}, and 
produces \emph{consistent execution results}.

\begin{figure*}[ht]
	\centering
	\includegraphics[width=\linewidth]{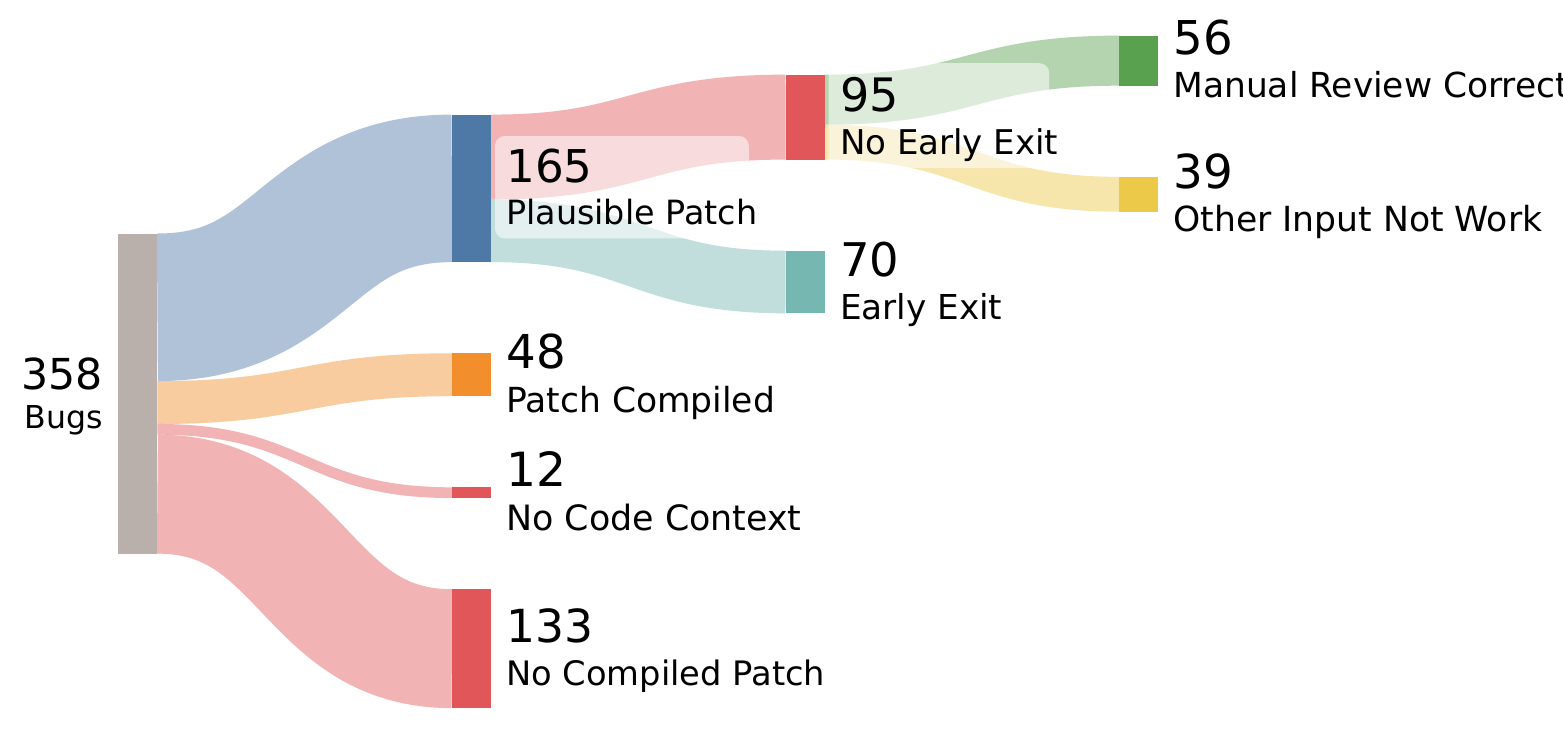}
	\caption{The actual fix ratio of \sys{-GPT-4o}.}
	\label{fig:appendix-actual-fix}
\end{figure*}

To better understand this gap, we analyze the patches generated 
by \sys{-GPT-4o} as an example, manually evaluating all \emph{plausible patches} 
as shown in~\autoref{fig:appendix-actual-fix}. Specifically, we first 
automatically validate whether the \emph{plausible patches} maintain the same 
execution path as the unpatched program when using the PoC as input. 
We then manually verify the patches to assess 
their effectiveness against a broader range of inputs.
We do not perform this verification for CodeRover-S or other 
state-of-the-art (SoTA) agents, as their patches often involve heavy 
modifications—such as code deletion or major rewrites—making automated 
validation infeasible.

\begin{figure}[t!] 
    \begin{lstlisting}[style=cpp]
// insert a counter 
printf("[SYS_INFO] hit crashsite for %d times\n", counter ++);
// optional: the patch may insert a check there.
stop_spatial(br->buffer, malloc_usable_size(br->buffer), 
            &br->buffer[cwords]);
// actual crash site 
b = br->buffer[cwords] << br->consumed_bits;
    \end{lstlisting}
    \caption{Example to automating the PoC execution comparision.}
    \label{list:appendix-validation}
\end{figure}

\noindent \textbf{Consistent Execution.} 
\sys only inserts checks without deleting or rewriting existing code. 
This patching strategy ensures that if the patched program does not 
exit prematurely, it preserves the original functionality, meaning it 
follows the same execution path as the unpatched program. Therefore, 
to validate the patch, we only need to verify whether the PoC reaches 
the crash site at the same point in both the patched and unpatched 
programs.
\autoref{list:appendix-validation} shows how we automate this PoC 
execution comparison by injecting a counter. We instrument both the 
patched and unpatched program source with the counter and compare the 
printed counter values in stdout. If the values are identical, it 
indicates that the patch does not introduce an early exit.
Due to the nature of \sys, the patch guarantees the \emph{consistent 
execution} as well.

\noindent \textbf{Manual Review.} 
However, this automated validation only ensures consistent behavior 
when using the PoC input. To further evaluate patch robustness, 
we manually review the patches that if their broader is correctly set, 
to determine their behaivor with a broader set of inputs.

\noindent \textbf{Successful Fixes in \sys.}
\autoref{fig:appendix-actual-fix} presents the results of our 
analysis. Among the 165 plausible fixes, only 95 patches avoid 
introducing early exits when tested with the PoC input. After 
manually inspecting these patches, we find that 56 of them also 
avoid early exits across different inputs, while the remaining 
39 may block some valid inputs (\ie \emph{Early Exit}). Nevertheless, 
all patches preserve program functionality as long as the program 
continues execution, which highlights a major advantage of \sys.

\begin{tcolorbox}[colback=gray!20, colframe=black, arc=3mm, boxrule=1.5pt]
    \textbf{Takeaway:} Developer need to manually verify the fix, even 
    its classified as plausible.
\end{tcolorbox}

\section{The Vulnerability Fixing Template}
\label{sec:append-fix-template}

\emph{Template-guided patch generation} is based on a set of predefined 
vulnerability templates. To construct these templates, we first analyze 
the distribution of bug types uncovered in real-world fuzzing 
campaigns, and subsequently select the top four categories to guide our 
template development.

\begin{figure*}[ht]
	\centering
	\includegraphics[width=\linewidth]{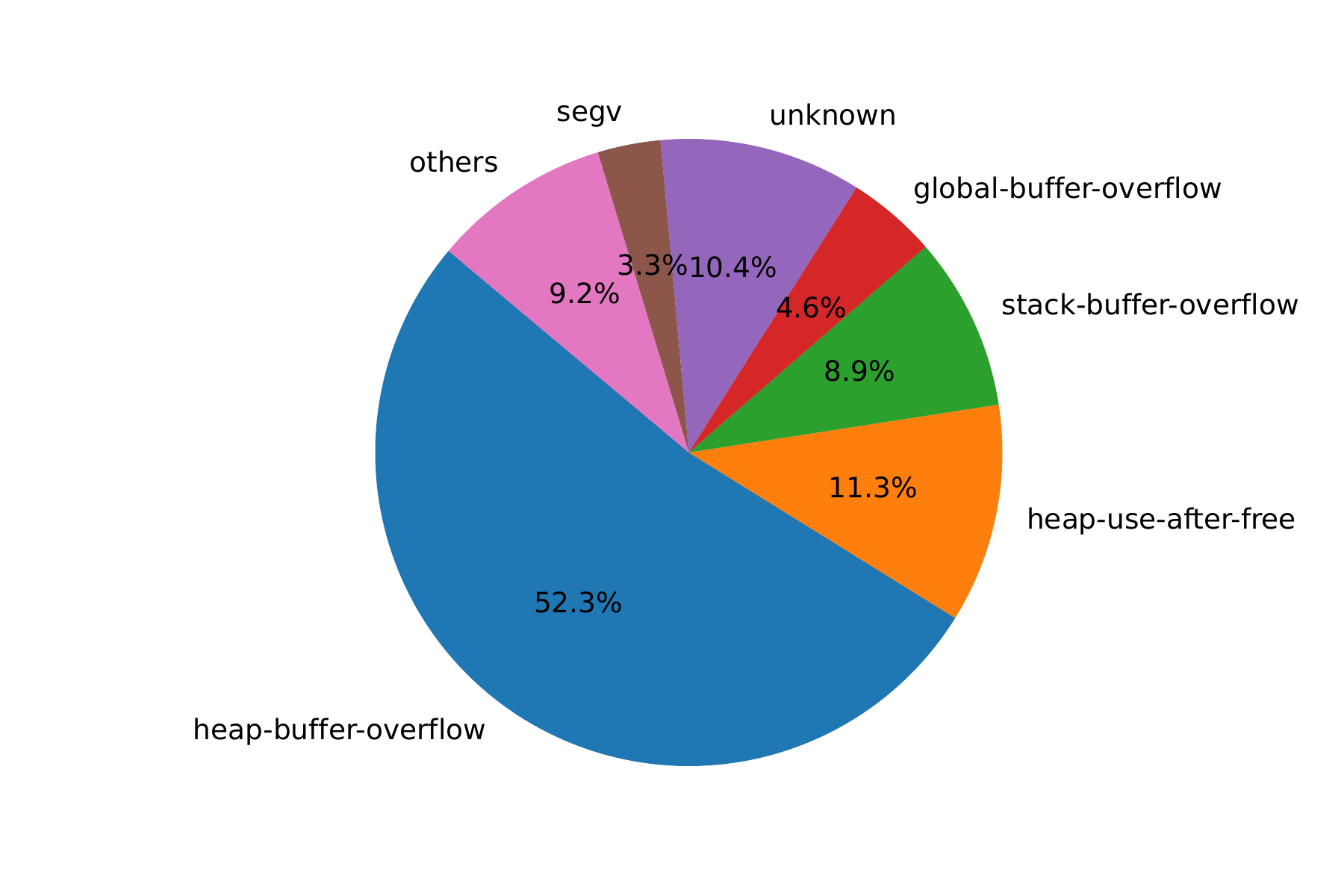}
	\caption{Exploitable Bug Types Distribution in ARVO. We merge the 
    bug categories less than 3\% into 'others'.}
	\label{fig:append-bug-distribute}
\end{figure*}

We begin by examining the distribution of bug types among real-world 
vulnerabilities discovered through OSS-Fuzz, using the ARVO 
dataset~\cite{mei2024arvo}. ARVO comprises over 5,000 bugs identified 
by Google's OSS-Fuzz~\cite{oss_fuzz}, providing a reliable and 
representative ground truth for our analysis. Our focus is on 
AddressSanitizer (ASan) bugs, as they remain the most exploited 
class of vulnerabilities in recent in-the-wild 0-day 
attacks~\cite{memory_corruption_exploitable}. 
\autoref{fig:append-bug-distribute} presents the distribution of ASan 
bug types within ARVO.

Overall, Heap Buffer Overflow (HBO) emerges as the most prevalent 
ASan vulnerability, accounting for 52.3\% of all ASan reports. 
Following this, Heap Use-After-Free (UAF) is the most common form of 
temporal memory corruption, representing 11.3\% of the reports. 
Stack Buffer Overflow (SBO) and Global Buffer Overflow (GBO) 
contribute 8.9\% and 4.6\% of the cases, respectively. We also 
observe instances labeled as "Unknown" or "Segv," often caused 
by out-of-bounds accesses that occur far from ASan redzone, making 
precise classification difficult.

Based on this analysis, we identify HBO, UAF, SBO, and GBO as the most 
representative ASan bug categories. Among them, HBO, SBO, and GBO are 
classified as spatial memory corruptions, while UAF represents a 
temporal memory corruption. Consequently, we develop templates that 
address both spatial and temporal memory error classes.

\begin{figure}[t!] 
    \begin{lstlisting}[style=cpp]
void stop_temporal(void *ptr) {
    // temporal memory corruption happens 
    if (is_destroyed(ptr)) {
        exit(0);
    }
}
// vulnerable code example
free(ptr);
... 
+ stop_temporal(ptr);
memcpy(dst, ptr, size);
\end{lstlisting}
\caption{Repairing Template For Temporal Memory Corruptions.}
\label{list:bug-fix-template-temporal}
\end{figure}

\noindent \textbf{Temporal Memory Corruption Template.} 
To mitigate \emph{temporal memory corruption}, the template 
\lstinline{stop_temporal} is introduced. This template takes a pointer 
currently in use by the program and checks whether it has already been 
deallocated. The \lstinline{is_destroyed} varies depending on the 
memory allocator used. \autoref{list:bug-fix-template-temporal} 
illustrates an example in which \sys inserts a call to 
\lstinline{stop_temporal(ptr)} immediately before the crash site, 
thereby preventing potential UAF exploitation.

\begin{figure}[t!] 
    \begin{lstlisting}[style=cpp]
void stop_spatial(void *buf, size_t buf_size, void * ptr) {
    // spatial memory corruption happens 
    if (ptr >= buffer + buf_size || ptr < buf) {
        exit(0);
    }
}
// vuln code example
for (int idx = 0; i < size; i ++) {
    // Global/Stack-Buffer-Overflow: crash site 
+   stop_spatial(buf, sizeof(buf), &buf[i]);
    // Heap-Buffer-Overflow: crash site 
+   stop_spatial(buf, malloc_usable_size(buf), &buf[i]);
    result_buf[i] = buf[i];
}
\end{lstlisting}
\caption{Repairing Template For Spatial Memory Corruptions.}
\label{list:bug-fix-template-spatial}
\end{figure}

\noindent \textbf{Spatial Memory Corruption Template.} 
To address \emph{spatial memory corruptions}, the corresponding 
template requires three arguments: \lstinline{buf}, the base address 
of the accessed buffer; \lstinline{buf_size}, the size of the buffer; 
and \lstinline{ptr}, the pointer performing the access that may be 
out-of-bounds. An example is shown in 
\autoref{list:bug-fix-template-spatial}. This template is applicable 
to SBO, GBO, and HBO. The main difference lies in how the buffer 
size is determined: for SBO and GBO, \lstinline{sizeof} is used, 
while for HBO, \lstinline{malloc_usable_size} is required to 
retrieve the dynamically allocated buffer's actual size.

\section{The Regex and Prompt Preparation}
\label{sec:append-regex-prompt}

This section introduce some technical details on how our 
\emph{regex-based context retrieval} works and how we prepare 
the LLM prompt for the \emph{crash site analysis}.

\begin{figure}[t!] 
\begin{lstlisting}[style=stacktrace]
==19==ERROR: AddressSanitizer: global-buffer-overflow on address 0x0000071da0a8 at pc 0x000001b03a92 bp 0x7ffdb1079ef0 sp 0x7ffdb1079ee8
READ of size 8 at 0x0000071da0a8 thread T0
SCARINESS: 33 (8-byte-read-global-buffer-overflow-far-from-bounds)
    #0 0x1b03a91 in wassp_match_strval /src/wireshark/epan/dissectors/packet-wassp.c:4384:32
    #1 0x1b03a91 in dissect_wassp_sub_tlv /src/wireshark/epan/dissectors/packet-wassp.c:4779
    #2 0x1b0348a in dissect_wassp_sub_tlv /src/wireshark/epan/dissectors/packet-wassp.c
    #3 0x1b06a96 in dissect_wassp_tlv /src/wireshark/epan/dissectors/packet-wassp.c
    #4 0x1b084b5 in dissect_unfragmented_wassp /src/wireshark/epan/dissectors/packet-wassp.c:5873:12
    #5 0x1b084b5 in dissect_wassp /src/wireshark/epan/dissectors/packet-wassp.c:6021
...    
\end{lstlisting}
\caption{AddressSanitizer Error Report for bug 20004.}
\label{fig:appendix-sanitizer-trace}
\end{figure}

\noindent \textbf{Regex-Based Context Retrieval.}
\sys takes AddressSanitizer error reports~\cite{serebryany2012addresssanitizer} 
as input, which follow a well-defined format~\cite{asan_callstack}. 
To process these reports, \sys implements a \emph{regex-based context 
retrieval} module. This module begins by extracting the pattern 
prefixed with \texttt{"AddressSanitizer: "} to determine the bug 
category. Depending on whether the bug is temporal or spatial in 
nature, \sys further extracts the crash and allocate call stack, 
omitting entries associated with library code (\eg frames containing 
\texttt{"compiler-rt"} in the file path). It then identifies the last 
stack frame corresponding to the user program and retrieves the 
associated source file and line number where the crash occurred. 
Finally, \sys extracts a window of code context, consisting of two  
lines before and after the crashing line in the identified source file.
\autoref{fig:appendix-sanitizer-trace} illustrates the stack trace of 
bug 20004. In this example, \sys correctly identifies the bug as a 
\texttt{global-buffer-overflow}, locates the crashing source file as 
\texttt{packet-wassp.c}, and extracts lines 4382 through 4385 as the 
surrounding context.

\begin{figure}[h!] 
\begin{lstlisting}[style=stacktrace]
You are a software developer maintaining a large project.
You are working on an issue submitted to your project.
The issue contains a description marked between <issue> and </issue>.
Another developer has already collected code context related to the issue for you.
Your task is to find potential root cause for the issue by answer the following 
questions.

The bug is a global-buffer-overflow and should be described in the following:
  - g_buffer: the global buffer that program trying to access, format is (void *)
  - g_buffer_size: the end of the global_buffer, presenting as sizeof($g_buffer)
  - g_ptr: the address of element visit when the program crash, format is (void *)

REMEMBER:
- Output in ```json ``` format.
- for g_ptr, ONLY USE VARIABLES THAT EXISTS IN <context> ... </context>
- for g_buffer, USE VARIABLES EITHER IN <context> ... </context> or <log> ... </log>
- for g_buffer, if <log_type> indicate its a struct, use (void *) ($g_buffer) to represent, else use (void *) ($g_buffer)
- DO NOT USE ANY INTERGER in the answer. 

For example, given the following input:
// example issue

You are suppose to return:
// example output

Here are your inputs:
<issue>
    <type>global-buffer-overflow</type>
    <crash_location>
    line 4384, column 32</crash_location>
    <log_type>array</log_type>
    <log>XXX is located 8 bytes to the right of global variable 'tlvMainTable'</log>
    <context>
    4373: static const char* wassp_match_strval(...)
    4374: {
    // ...
    4382: 	}
    4383: 
    4384: 	return in_ptr->entry[in_type].name;
    4385: }
    4385: }
    </context>
</issue>
\end{lstlisting}
\caption{LLM Prompt to analyze bug 20004.}
\label{fig:appendix-analysis-prompt}
\end{figure}

\noindent \textbf{Prompt Preparation.}
Building on the retrieved context information, \sys performs 
\emph{crash-site analysis} by leveraging a predefined prompt in 
conjunction with the contextual code. Specifically, as illustrated 
in \autoref{fig:appendix-analysis-prompt}, \sys first issues a 
structured prompt and then embeds the extracted context within 
an \lstinline{<issue>} tag to facilitate LLM analysis.

\end{document}